\documentclass[11pt]{article}
\usepackage[utf8]{inputenc}
\usepackage[fleqn]{amsmath}
\usepackage{amsmath,amsfonts}
\usepackage[english]{babel}
\usepackage{textcomp}
\usepackage{caption,subcaption}
\usepackage{graphicx}
\usepackage{rotating}
\usepackage{authblk}
\usepackage{bookmark}
\usepackage{float}
\usepackage{amsmath}
\usepackage[square,numbers]{natbib}
\usepackage{color, soul}
\hypersetup{pdfauthor={some author},pdftitle={eye-catching title}}
\usepackage{ifpdf}

\title{\textbf{Gravitationally collapsing stars in $f(R)$ gravity}}
\author{{Suresh C. Jaryal \footnote{suresh.fifthd@gmail.com}}}
\author{{Ayan Chatterjee\footnote{ayan.theory@gmail.com}}}
\affil{{Department of Physics and Astronomical Science }\\
{ Central University of Himachal Pradesh\\
Dharamshala, Kangra (HP), India 176215}.}
\date{}
\begin{document}
\maketitle
The gravitational dynamics of a collapsing matter configuration which is simultaneously
radiating heat flux is studied in $f(R)$ gravity. Three particular functional forms in $f(R)$ gravity are considered to show that it is possible to envisage boundary conditions such that the end state of the collapse has a weak singularity and that
the matter configuration radiates away all of its mass before collapsing to reach the central singularity.
\\\\
{\bf{Keywords:}}\,\,Gravitational collapse. $f(R)$ gravity.
\section{Introduction}
The general theory of relativity (GR) is an impeccably robust
relativistic theory of gravity.
It forms the basis for our understanding of gravitational phenomenon at
small as well as large scales
\cite{Hawking1975, Wald}.
However, it is well known that GR cannot be the ultimate theory of
gravitation since it
has a well defined regime of validity; for example, understanding past
and future spacetime
singularities are beyond the reach of GR.
Suitable modification(s) of GR is(are) essential to understand
singularities,
or to resolve them. It is believed that a quantum theory of gravity may
lead to solution
to the problems affecting GR \cite{Wald, Joshi, Wald_qft, buchbinder,
	reuter_book}.
Naturally, in absence of any consensus on the theory of quantum gravity,
modified
theories of gravity with quantum corrections are also of interest.
The Einstein- Hilbert action may be thought of as only a low energy
contribution
and higher curvature terms consistent with the diffeomorphism invariance
may become
relevant as one goes to higher energies. Higher curvature corrections
should leave imprints
at low energy scales which become important for low energy physics
too \cite{buchbinder, reuter_book, green_schwartz_witten}.
Out of these alternate theories, we shall study the $f(R)$ model
since it has been found interesting in the cosmological studies as well.
These $f(R)$ gravity models are thought of as alternate
to the dark energy models \cite{SC_VFC_SC_AT, Nojiri:2003ft,
	SMC_VD_MT_MST}.
The standard way to construct these $f(R)$ theories
is to replace the Einstein- Hilbert Lagrangian
by a well defined function of Ricci scalar, $f(R)$
(for general relativity $f(R)=R$) \cite{HAB_JDB}.
For a detailed review of
the motivation, validity of various functional forms $f(R)$,
applications as well as shortcomings of
these gravity theories have
been extensively analyzed \cite{Kobayashi:2008tq,Sotiriou:2008rp, TC_PGF_AP_CS, TPS_VF,
	Capozziello:2011nr, Capozziello:2011gm, Astashenok:2013vza, SN_SDO,
	SN_SDO_VKO}.

The purpose of the present work is to construct some examples of
spacetimes in $f(R)$ gravity
which admit gravitationally weak spacetime singularities \cite{Clarke:1994cw}.
The physical situation which we consider here is
the following: the initial mass of
the collapsing star or matter configuration is so high
that the force of gravity overwhelms the thermal or quantum mechanical
pressures. As a result, no stable configuration such as a neutron star
or a white dwarf exists during the collapse process.
Generically, such collapsing matter configurations admit
diverging density and curvatures at
the central singularity (we must emphasize that
spherically symmetry is assumed throughout the collapse). However, we
shall show below that
it is possible, without violating energy conditions, that
during the gravitational collapse matter is radiated away at such a rate
that
the matter boundary never reaches its horizon. In this case,
no horizon is formed at the boundary since the star radiates off all its
mass before reaching the singularity. The central singularity is naked
but is gravitationally weak.
During this process, all the physical quantities energy density, radial and
tangential pressure, pressure anisotropy, heat flux, remain regular and
positive throughout the collapse. The luminosity and adiabatic index are
also
regular and positive,
and admits maximum value when the matter approaches the singularity. Thus,
for an observer at infinity observing the collapse, the configuration
shall become extremely bright, reaching its
maximum luminosity before turning off, indicating that it has radiated
off all its mass.
Solutions of such kind are not unknown and is possible in the Newtonian
gravity as well.
Consider a star in the Newtonian gravity which is extremely heavy to be
supported by
the Pauli exclusion principle alone. So, when the gravitation
contraction takes place, thermal pressure
may balance to some extent. But since there is no event horizon, the
star shall continue to radiate
all the gravitational energy to infinity. Hence all the matter contained
in the star shall
be converted to thermal radiation and radiated off.
We show here that configurations of similar nature are also
possible in
$f(R)$ gravity.

From a theoretical point of view,
gravitational collapse of matter is important and
recently, there has been an increasing interest to understand
whether the nature of collapse is altered in modified gravity \cite{Astashenok:2013vza,DeFelice:2010aj,Babichev:2009td,Cooney:2009rr,Astashenok:2017dpo}.
Although it
remains to study in detail if
compact objects formed in theories like the $f(R)$ gravity can be
experimentally detected,
several investigations have already been carried out in this direction .
The idea is to use the multi-wavelength and well as the gravitational
wave data to probe matter under extreme gravity (like
the composition of the inner core of a neutron star) which remain
unknown, as neither they can they be probed directly with astrophysical
observations nor
are they they within the reach of the present- day theory experiments.
Alternatively, the direct gravitational wave data have also made it possible
to investigate if these objects are possible in alternate gravity theories.
In particular, the recent detection of GW190814, have fueled the speculation
that it might be the heaviest neutron star till date \cite{Abbott:2020khf}. It has been argued
that some models of $f(R)$ gravity can indeed explain the origin of such a
large mass of $(2.5$-$2.7)M_{\odot}$, even using the well- known equations
of state \cite{Astashenok:2020qds,Astashenok:2021peo}.

In this paper, we shall not concern ourselves 
with compact objects like the white dwarf or a neutron star.
We shall assume that the matter is extremely massive, and such objects
continue to collapse under its own gravity.
The gravitational collapse phenomenon in GR show that the collapse outcome
depends upon, among other quantities, the choices of mass profiles and
velocity profiles of the collapsing matter. In the context of
inhomogeneous LTB models in GR,
these issues have been considered in great detail for various matter
models including dust and
viscous fluids \cite{Chatterjee:2020khj}.
The stellar collapse of stellar collapse in $f(R)$ gravity
using different forms of $f(R)$ function and different
matter distributions may be
found in\,\cite{KB_SN_SDO,EVA_ADD,MS_HRK,RG_AMN_SDM_SGH,SC_NBa,SC_NBb,Astashenok:2018bol,Abbas:2017kra,SC_RG_SDM_NB,GA_HN,MS_ZYa,MS_ZYb,HRK_IN,AB_BJ_PZ,JG_DW_AVF}.
Although different type of $f(R)$ models may be considered, only the ones
which are in agreement with the standard cosmological observations
should be of interest. Here, we consider three
models of $f(R)$ gravity given respectively by: $(a)~f(R)=(R+\lambda R^{2})$ \cite{AAS1980}\, ,\,
$(b) ~f(R)\sim R^{1+\epsilon}$, with $\epsilon <<1$ \cite{SC_NBb} and $(c)~f(R)=R+\lambda\,\left[
\exp{(-\sigma R)}-1\right]$, with $\lambda\sigma <1$,  \,\cite{Cognola:2007zu}. 
All three theories play a major role in the study of the early and the late universe.
For example, in the $R+\lambda R^2$ model, 
the $R^{2}$ term is weak in the weak gravity regime.
But it also contributes significantly during the early 
universe and in strong gravity regime. Here, we
study gravitational collapse of an extremely massive object (so massive that its 
gravitational attraction predominates thermal or quantum mechanical pressures), and so naturally,
since we are in a strong gravitational field, this 
model may assume significance.  
In all these theories, we shall show that an extremely
massive spherical matter cloud, admitting radial and tangential pressures, and outgoing
heat flux, can collapse
in a manner that no horizon is formed at the boundary (since the star
radiates off all its
mass before reaching the singularity) making the central singularity
naked but weakly singular. To ensure that the gravitational
collapse proceeds continuously without forming any stable object, 
we assume the radius of the $2$- sphere cross- sections decrease linearly with time.
The interior collapsing spacetime shall be smoothly matched with the
exterior
Vaidya spacetime\,\cite{Vaidya} over a timelike surface
$\Sigma$ \cite{Santos}.

We approach the problem as follows.
In section \ref{sec2}, we give the field equations of $f(R)$ gravity and
the
junction conditions for smooth matching of the interior and
the exterior spacetimes across the timelike hypersurface $\Sigma$.
This section also includes the solutions of the $f(R)$ field equations
along with
the explicit expressions
for physical quantities. For the solution, we use
the Karmarkar condition \cite{Eisenhart1925a,Eisenhart1925b, Karmarkar}.
These conditions determine
gravitational potentials for static and non- static systems
\cite{SP2016,NGM2018,Govender:2020gyc, SCJ}.
We must mention here that similar studies on $f(R)$ gravity have
been carried out in \cite{GA_HN}. However,
the solutions obtained there are restrictive in the sense that one of
the metric function have been kept constant to derive the values of
other metric function. On
the other hand we shall show that such conditions are overly restrictive.
Note that since Karmarkar condition expresses relationship between
metric functions, the forms of metric functions are arbitrary, and
dependent on
one's choice. However we argue that this arbitrariness
may be removed if metric functions are related
to matter variables. For example, if we assume a
specific form of pressure anisotropy (difference of the radial and
tangential pressures, denoted by $\Delta=p_{t}-p_{r}$), this
gives rise to unambiguous set of gravitational potentials, and a class
of metric representing collapsing spacetimes. We inspect
the physical relevance of these exact solutions by verifying the energy
conditions. The stability criteria and discussion about the luminosity and
adiabatic index, radial and transverse velocity are carried out in
section \ref{sec3}. It is shown that a faraway observer will see a source
whose luminosity is exponentially increasing until a time
when it shuts off quickly. This is due to the fact that
the total mass of the star radiates linearly and, as the star reaches
its maximum luminosity there is no mass left to radiate. The evolution
of the temperature profiles during stellar collapse is also studied in
section \ref{sec5}
since they play an important role in the study of transport processes
in radiative gravitational collapse \cite{Maartens1995,Martinez1996,Herrera1996,Herrera1997,LHerrera,GMM1998,GMM1999,GGM1999,GG2001}.
In section \ref{sec6} contains discussion of the results
accompanied with concluding remarks.
\section{Field Equations and Matching Conditions}\label{sec2}
The action for the $f(R)$ gravity is obtained by replacing the standard 
Einstein- Hilbert Lagrangian by a well defined function of Ricci scalar \cite{TPS_VF}
\begin{eqnarray}
\mathcal{S}&=&\frac{1}{2}\,\int \sqrt{-g}\,\left[\,f(R)+2\, \mathcal{L}_{M}(g_{\mu\nu},\Psi_{m})\right]d^{4}x\, , \label{action} 
\end{eqnarray}
where $\Psi_{m}$ refers collectively to all matter fields, 
$ \mathcal{L}_{M}$ is the Lagrangian density of the matter fields $\Psi_{m}$,
$g$ is the determinant of the metric tensor $g_{\mu\nu}$,
$R$ is the Ricci scalar curvature and $f(R)$ is the generic function of Ricci scalar
defining the theory under consideration and $($using units with $c=1=8\pi G)$.
Varying the action \eqref{action} with respect to the metric tensor $g_{\mu\nu}$
yields the following field equations:
\begin{eqnarray}
F(R)\,R_{\mu\nu}-\frac{1}{2}f(R)\,g_{\mu\nu}-\left(\nabla_{\mu}\,\nabla_{\nu}-g_{\mu\nu}\,\Box \right)F(R)=T^{M}_{\mu\nu},
\end{eqnarray}
where $F(R)={d\,f(R)}/{dR}$,\, and $\Box\equiv\nabla_{\mu} \nabla^{\mu}$.
This equation may also be rewritten as
\begin{eqnarray}
R_{\mu\nu}-(1/2)\, g_{\mu\nu}\,R=F(R)^{-1}\,\left(T^{M}_{\mu\nu}+T^{D}_{\mu\nu}\right)\, ,\label{Var_action}
\end{eqnarray}
where the left side of the equation \eqref{Var_action} is the usual Einstein tensor, $T^{M}_{\mu\nu}$ 
and $T^{D}_{\mu\nu}$ are the energy momentum tensor and effective energy momentum tensor having the form as:
\begin{eqnarray}
T^{M}_{\mu\nu}&=&(p_{t}+\rho)u_\mu u_\nu +p_{t} g_{\mu\nu}+(p_{r}-p_{t})X_{\mu}X_{\nu}+q_{\mu}u_{\nu}+q_{\nu}u_{\mu}\, , 
\label{tmnz}\\
T^{D}_{\mu\nu}&=&(1/2)\left[f(R)-R\,F(R)\right]\, g_{\mu\nu}+\left(\nabla_{\mu}\nabla_{\nu}-g_{\mu\nu}\,\Box \right)F(R).
\label{darktmnz}
\end{eqnarray}
Here, $\rho$, $p_{r}$ and $p_{t}$ are the energy density, radial pressure and 
the tangential pressure respectively.
Also, $q^{\mu}$
$u^{\mu}$, $X^{\mu}$ represents the radial heat flow vector, $4$-velocity vector and spacelike $4$-vector respectively, which satisfy $u_\mu u^\mu=-X_\mu X^\mu=-1$ and $u_\mu X^\mu=u_\mu q^\mu=0$.

We now consider a general non- static shear free spherically symmetric spacetime
metric given by the following form 
\begin{equation}
 ds^{2}=-a(r)^2 dt^2 + b(r)^2 s(t)^2 \left(dr^2 + r^2 d \theta^2+ r^{2} \sin^2{\theta} d\phi^2 \right).
\label{1eq1}
\end{equation}
The forms of $u^{\mu}$, $X^{\mu}$ and $q^{\mu}$ in terms of the metric \eqref{1eq1} are 
\begin{equation}
 u^\mu=a^{-1}\,\delta^{\mu}_{0}\,; ~~~~\,\,\,\, X^{\mu}=(b\,s)^{-1}\,\delta^{\mu}_{1}\,\,;~~~\,\,\,\,
 q^{\mu}=(b\,s)^{-1}\,X^{\mu} \label{uXq},
\end{equation}
The magnitude of the expansion scalar $\Theta$ and Ricci scalar for the metric \eqref{1eq1} have the form
\begin{eqnarray}
\Theta&=&\bigtriangledown_{\mu}u^{\nu}=\frac{3\,\dot{s}}{a\,s},\label{Theta}\\
R&=&6\,\frac{s\ddot{s}+\dot{s}^{2}}{a^2\,s^2}-\frac{2}{b^2\,s^2}\left[\frac{a^{\prime\prime}}{a}-
\frac{b^{\prime 2}}{b^{2}}+\frac{a^{\prime}b^{\prime}}{a\,b}
+2\frac{b^{\prime\prime}}{b}+\frac{2}{r}\left(\frac{a^{\prime}}{a}+2\frac{b^{\prime}}{b} \right)\right].
\label{RScalar}
\end{eqnarray}
The field equations in $f(R)$ gravity for the 
 metric \eqref{1eq1}, energy momentum tensor \eqref{tmnz}, \eqref{darktmnz} and \eqref{uXq} are  
 \begin{equation}
\rho=\frac{F}{s^2}\left[\frac{3\,\dot{s}^{2}}{a^2}-\frac{1}{b ^{2}}\left(\frac{2\,b^{\prime\prime}}{b}
-\frac{b^{\prime}\,^{2}}{b\,^{2}}+\frac{4}{r}\frac{b^{\prime}}{b} \right)\right]
+\left(\frac{f-R\,F}{2}\right)+\frac{3\dot{s}\dot{F}}{s\,a^{2}}-\frac{1}{b^{2}\,s^{2}}\left[F^{\prime\prime}
+F^{\prime}\left(\frac{b^{\prime}}{b}+\frac{2}{r}\right)\right]\label{rho},
\end{equation}
 \begin{eqnarray}
{p}_{r}&=&\frac{F}{s^2}\left[-\frac{1}{a^{2}}\left(2s\,\ddot{s}+\dot{s}^{2}\right)
+\frac{1}{b^{2}}\left(\frac{2\,a^{\prime}\,b^{\prime}}{a\,b}+\frac{2}{r}\left(\frac{a^{\prime}}{a}
+\frac{b^{\prime}}{b}\right)+\frac{b^{\prime}\,^{2}}{b\,^{2}} \right)\right]
-\left(\frac{f-R\,F}{2}\right)\nonumber\\
&&~~~~~~~~~~~~~~~~~~~~~~~~~~~~~~~~~~~~-\frac{\dot{F}}{a^2}\left(\frac{\ddot{F}}{\dot{F}}+\frac{2\dot{s}}{s}\right)+\frac{F^{\prime}}{b^{2}\,s^{2}}\left(\frac{a^{\prime}}{a}+\frac{2}{r}+\frac{2b^{\prime}}{b} \right)\label{pr},
\end{eqnarray}
 \begin{eqnarray}
{p}_{t}&=&\frac{F}{s^2}\left[-\frac{1}{a^{2}}\left(2s\,\ddot{s}+\dot{s}^{2}\right)
 +\frac{1}{b^{2}}\left(\frac{a^{\prime\prime}}{a}+\frac{b^{\prime\prime}}{b}-\frac{b^{\prime}\,^{2}}{b\,^{2}}+\frac{1}{r}\left(\frac{a^{\prime}}{a}+\frac{b^{\prime}}{b}\right)
  \right)\right]
 -\left(\frac{f-R\,F}{2}\right)\nonumber\\
&& ~~~~~~~~~~~~~~~~~~~~~~~~~~~~~~~~~
-\frac{\dot{F}}{a^2}\left(\frac{\ddot{F}}{\dot{F}}
 +\frac{2\dot{s}}{s}\right)+\frac{1}{b^{2}\,s^{2}}\left(F^{\prime \prime}
 +F^{\prime}\left(\frac{a^{\prime}}{a}+\frac{1}{r}\right)\right)
 \label{pt},
 \end{eqnarray}
 \begin{equation}
q=-\frac{2\,a^{\prime}\,\dot{s}\,\dot{F}}{a^{2}\,b^{2}\,s^{3}}
 +\frac{1}{a^{2}\,b^{2}\,s^{2}} \left(\dot{F}\,^{\prime}-\frac{\dot{F} a^{\prime}}{a}-\frac{\dot{s} \, F^{\prime}}{s} \right)\label{q},
 \end{equation}
where prime and dot are the derivatives with respect to $r$ and $t$ respectively.

Let us consider the junction conditions
for the smooth matching of the interior manifold $\mathcal{M^-}$ (equation \eqref{1eq1} considered above)
with the exterior manifold $\mathcal{M^+}$ across timelike hypersurface $\Sigma$,
at  $r=r_b$. As described in \cite{Deruelle:2007pt,Senovilla:2013vra}, the junction conditions for the $f(R)$ gravity 
requires the matching of several geometric quantities other than
the induced metric ($h_{ij}$) and the extrinsic curvature ($K_{ij})$. 
In fact, it has been established that in
$f(R)$ gravity, the following variables must be matched at the boundary:
\begin{eqnarray}\label{junction}
[h_{ij}]_{-}^{+}&=&0, \\
F(R)\,[K_{ij}-(1/3)K\, h_{ij}]_{-}^{+}&=&0, \\
~[K]_{-}^{+}&=&0, \\
(\partial F(R)/ \partial R)\,[\partial_{\tau}R]_{-}^{+}&=&0, \\
~[R]_{-}^{+}&=&0,
\end{eqnarray}
where $\tau$ represents the proper time of the timelike hypersurface, and $K$ is the trace of the
extrinsic curvature. 
Out of these five conditions, for the set of $f(R)$ theories under consideration,
it is sufficient to match the metric, the extrinsic curvature, the Ricci scalar,
and the derivative of the Ricci as given above, determined from either sides.    
\\
\\
The Vaidya spacetime in the outgoing coordinate is taken to be our exterior spacetime $\mathcal{M^+}$
\cite{Vaidya}
\begin{eqnarray}
ds^{2}_{+}=-\left[1-\frac{2\,M(v)}{{\bf{r}}}\right]\, dv^{2}-2dvd{\bf{r}}+ {\bf{r}}^2 \left(d \theta^2
 +\sin^2\theta d\phi^2\right)\label{m+},
\end{eqnarray}
 For our later convenience, let us define the proper time, $d\tau=a(r)_{_{\Sigma}}\, dt $.  
The junction condition as given \eqref{junction}, implies the following conditions
\begin{eqnarray}
{\bf{r}}_{_{\Sigma}}(v)&=&(r\,b\,s)_{_{\Sigma}} ,\label{rR}\\
\left(\frac{dv}{d\tau}\right)_{\Sigma}^{-2}&=&\left( 1-\frac{2M}{\bf{r}}+2\frac{d\bf{r}}{dv}\right)_{\Sigma}, \label{vtau}
\end{eqnarray}
where $\tau$ represents the proper time defined on the hypersurface $\Sigma$. The normal vector fields to $\Sigma$ are given by
\begin{equation}
n^{-}_{l}=\left[0,(b\,s)_{_{\Sigma}},0,0\right], ~~~~~~~
n^{+}_{l}=\left[1-\frac{2M}{{\bf{r}}}+2\frac{d{\bf{r}}}{dv}\right]^{-\frac{1}{2}}_\Sigma 
\left[-\frac{d{\bf{r}}}{dv}\delta^0_l+\delta^1_l\right]_{\Sigma} .
\end{equation}
The extrinsic curvatures for metrics \eqref{1eq1} and \eqref{m+} are given by
\begin{eqnarray}
K^{-}_{\tau\tau}&=&-\left[\frac{a^{\prime}}{a\,b\,s}\right]_{\Sigma},
K^{-}_{\theta\theta}=\left[r\,b\,s\left(1+\frac{r\,b^{\prime}}{b} \right)\right]_\Sigma , \\
K^{+}_{\tau\tau}&=&\left[\frac{d^2v}{d\tau^2}\left(\frac{dv}{d\tau}\right)^{-1}-\left(\frac{dv}{d\tau}\right)
\frac{M}{{\bf{r}}^2}\right]_{\Sigma},
K^{+}_{\theta\theta}=\left[\left(\frac{dv}{d\tau}\right)\left(1-\frac{2M}{{\bf{r}}}\right)
{\bf{r}}-{\bf{r}}\frac{d{\bf{r}}}{d\tau}\right]_\Sigma ,\label{Kthth+}\\
K^{-}_{\phi\phi}&=& \sin^2{\theta} K^{-}_{\theta\theta}\,\, , \,\, K^{+}_{\phi\phi}=\sin^2{\theta}
 K^{+}_{\theta\theta} .
\end{eqnarray}
Now, from the junction condition on $K_{ij}$ (because of the conditions that $K$ must satisfy
on the hypersurface, matching $K_{ij}$ is enough), we get the following.
From the equality for the $\theta\theta$ components
 at hypersurface $\Sigma$, and the equations \eqref{rR} and \eqref{vtau}, we obtain
\begin{eqnarray}
\left[r\,b\,s\left(1+\frac{r\,b^{\prime}}{b} \right)\right]_\Sigma&=&
\left[\left(\frac{dv}{d\tau}\right)\left(1-\frac{2M}{{\bf{r}}}\right)
{\bf{r}}-{\bf{r}}\frac{d{\bf{r}}}{d\tau}\right]_\Sigma\label{Kth+-},
\end{eqnarray}
and the total energy inside the boundary hypersurface $\Sigma$, given by 
the Misner- Sharp mass, denoted by $2m$ (such that $m=M$ on the matching hypersurface) \cite{Misner-Sharp,CM}, where
\begin{eqnarray}
m_{\Sigma}&=&\left[\frac{r^{3}\,\dot{s}^{2}\,b^{3}s}{2\,a^{2}}-\frac{r^{3}\,s\,b^{\prime}\,^{2}}{2\,b}
-r^2\,s\,b^{\prime}\right]_{\Sigma}. \label{mass}
\end{eqnarray}
Now, again from the matching of the $K^{+}_{\tau\tau}=K^{-}_{\tau\tau}$ 
component we have the following equation:
\begin{eqnarray}
-\left[\frac{a^{\prime}}{a\,b\,s}\right]_{\Sigma}&=&\left[\frac{d^2v}
{d\tau^2}\left(\frac{dv}{d\tau}\right)^{-1}-\left(\frac{dv}{d\tau}\right)\frac{M}{{\bf{r}}^2}\right]_\Sigma, \label{Ktt+-}
\end{eqnarray}
and, substituting the relation between proper and coordinate time along with 
the eqns. \eqref{rR} and \eqref{mass} into the eqn.\eqref{Kth+-} we have 
\begin{eqnarray}
\left(\frac{dv}{d\tau}\right)_{\Sigma}&=&\left(1+\frac{r\,b^{\prime}}{b}+\frac{r\,b\,\dot{s}}{a}\right)_{\Sigma}^{-1}. \label{dvdtau}
\end{eqnarray}
Now, differentiating \eqref{dvdtau} with respect to the $\tau$ and using eqns 
\eqref{mass} and \eqref{dvdtau}, we can rewrite \eqref{Ktt+-}. Further, 
comparing with equations \eqref{pr} and \eqref{q} we have the following useful form
\begin{eqnarray}
\left(p_{r}+T^{D}_{rr}+b\,sT^{D}_{tr}\right)_{_{\Sigma}}&=&(q\,b\,s)_{_{\Sigma}}. \label{matching}
\end{eqnarray}
where,
\begin{eqnarray}
T^{D}_{rr}&=&\left(\frac{f-R\,F}{2}\right)+\frac{\dot{F}}{a^2}\left(\frac{\ddot{F}}{\dot{F}}+\frac{2\dot{s}}{s}\right)
-\frac{F^{\prime}}{b^{2}\,s^{2}}\left(\frac{a^{\prime}}{a}+\frac{2}{r}+\frac{2b^{\prime}}{b} \right),\label{darkT_rr}\\
T^{D}_{tr}&=&\frac{1}{a^{2}\,b^{2}\,s^{2}} \left(\dot{F}\,^{\prime}-\frac{\dot{F}a^{\prime}}{a}-\frac{\dot{s}\,F^{\prime}}{s} \right),
\label{darkT_tr}
\end{eqnarray}
are the dark source terms. From equation \eqref{matching}, it is found that just 
like for general relativity, the radial pressure does not
vanish at the boundary but, instead is proportional to 
the dissipative as well as radiative dark source terms.
The extra terms $T^{D}_{rr}$ and $T^{D}_{tr}$ on the LHS of 
equation \eqref{matching} are the dark source term and
may appear due to the higher order curvature geometry of the collapsing sphere \citep{GA_HN}.

Let us now move to match $K$ as given in \eqref{junction}.
The expressions for the trace of extrinsic curvatures on the either sides lead to the 
following matching condition on the hypersurface:
\begin{eqnarray}
\left[p_{r}+T^{D}_{rr}+b\,sT^{D}_{tr} -q\,b\,s\right]_{_{\Sigma}} = 2\,[M - m]_{_{\Sigma}},
\end{eqnarray}
and naturally, this condition is identically satisfied due to the abovementioned equations. The matching
of the Ricci and its proper time derivative gives the following conditions which are to be satisfied
for the metric of the internal manifold (at the hypersurface $\Sigma$):
\begin{eqnarray}\label{Ricci_junction}
s\ddot{s}+\dot{s}^{2}&=&\frac{a^2}{3b^2}\left[\frac{a^{\prime\prime}}{a}+2\frac{b^{\prime\prime}}{b}+\frac{a^{\prime}b^{\prime}}{a\,b}+\frac{2}{r}\left(\frac{a^{\prime}}{a}+2\frac{b^{\prime}}{b} \right)-
\frac{b^{\prime 2}}{b^{2}}\right],\nonumber \\
3\dot{s}\ddot{s}+s\dddot{s}&=&0.
\end{eqnarray}
The metric of the internal manifold must be chosen so as to satisfy the two conditions
in \eqref{Ricci_junction}. To determine metric functions according to all 
these junction conditions, it is necessary to use some auxiliary conditions.
We shall see below that these equations are consistent with a collapsing time dependent
internal metric. In fact, one may argue that junction conditions indeed force
such a possibility.  
Additionally, we must also ascertain the physical viability of the spacetime metric.
From equations \eqref{pr}, \eqref{pt} and \eqref{1eq1}, 
the pressure anisotropy factor $\Delta=p_{t}-p_{r}$ has the form
\begin{eqnarray}
\Delta&=&\frac{F}{b^2\,s^2}\left[\frac{b^{\prime\prime}}{b}-\frac{2\,b^{\prime\,^{2}}}{b^2}+\frac{a^{\prime\prime}}{a}
-\frac{2\,a^{\prime}\,b^{\prime}}{a\,b}-\frac{1}{r}\left(\frac{a^{\prime}}{a}+\frac{b^{\prime}}{b}\right)
\right]+\frac{F^{\prime\prime}}{b^2\,s^2}-\frac{F^{\prime}}{b^2\,s^2}\left(\frac{2b^{\prime}}{b}+\frac{1}{r} \right). \label{Delta}
\end{eqnarray}
The general expression for the shear free spacetime as given in \eqref{Delta} is has the complicated form.
To find the solution of the metric functions and mathematical simplicity, we take an adhoc form of the pressure anisotropy $\Delta$ to be:
\begin{eqnarray}
\Delta &=& \frac{F^{\prime\prime}}{b^2\,s^2}-\frac{F^{\prime}}{b^2\,s^2}\left(\frac{2b^{\prime}}{b}+\frac{1}{r}\right)-\frac{F}{b^2\,s^2}\left[\frac{2\,a^{\prime}\,b^{\prime}}{a\,b}-\frac{a^{\prime\prime}}{a}+\frac{a^{\prime}}{r\,a}\right]\label{Delta1}
\end{eqnarray}
Although, we have chosen this form of the anisotropy in pressure $\Delta$ 
for the mathematical simplicity, later we will see that 
they represents the physically viable solutions of the potentials. Also, this choice of $\Delta$ is physically significant,
such that $\Delta$ is regular throughout the collapse.
It must be noted that this choice of the anisotropy $\eqref{Delta1}$
reduces the total pressure anisotropic equation \eqref{Delta} as differential equation of
only one function, given by
\begin{eqnarray}
0&=&\frac{1}{s^2\,b^2}\left(\frac{b^{\prime\prime}}{b}-\frac{2\,b^{\prime\,^{2}}}{b^2}-\frac{b^{\prime}}{r\,b}  \right),
\end{eqnarray}
The form of the function $b(r)$ is
\begin{eqnarray}
b(r)&=&-2[C_3\,r^2+2\,C_4]^{-1}\label{br},
\end{eqnarray}
where $C_3$ and $C_4$ are constant of integration.

Let us now use the fact that under certain conditions, a $(n+1)$-dimensional space can be
 embedded into a pseudo Euclidean space of dimension $(n+2)$ \citep{Eisenhart1925a}. Thus the necessary and sufficient condition for any Riemannian space to 
be an embedding class $\bf{I}$ is the Karmarkar condition \cite{Eisenhart1925b, Karmarkar},
%
\begin{eqnarray}
R_{rtrt}\,R_{\theta\phi\theta\phi}=R_{r\theta r\theta}\,R_{t\phi t\phi}
-R_{\theta rt\theta}\,R_{\phi rt\phi}.\label{Karmarkar}
\end{eqnarray}
%
The non vanishing components of the Riemann tensor for the metric \eqref{1eq1} are 
\begin{eqnarray}
R_{rtrt}&=& a^{2}\left(\frac{a^{\prime\prime}}{a}-\frac{b^{2}s}{a^{2}}\ddot{s}-\frac{a^{\prime}}{a}\frac{b^{\prime}}{b} \right)\,,\,\,\, R_{\theta\phi\theta\phi}={r^{4}b^{2}s^{2}}\left(\frac{b^{2}}{a^{2}}\dot{s}^{2}-\frac{2b^{\prime}}{rb}
-\frac{b^{\prime\,^{2}}}{b^{2}} \right)\sin^2\theta, \label{Rrtrt}\\
R_{r\theta r\theta}&=&{r^2b^{2}s^{2}}\left(\frac{b^{2}}{a^{2}}\dot{s}^{2}-\frac{b^{\prime}}{rb}
-\frac{b^{\prime\prime}}{b}+\frac{b^{\prime\,^{2}}}{b^{2}} \right)\,,\,\,\, 
R_{\theta rt\theta}=\frac{r^{2}b^{2}s}{a}a^{\prime}\dot{s}, \label{Rrthrth}\\
R_{t\phi t \phi}&=&{r^2a^2b}\left(\frac{a^{\prime}}{ra}-\frac{b^2s}{a^2}\ddot{s}
+\frac{a^{\prime}}{a}\frac{b^{\prime}}{b} \right)\sin^2\theta \,,\,\,\,
R_{\phi rt\phi}=\sin^2\theta R_{\theta rt\theta} .\label{Rphrtph}
\end{eqnarray}
Using the expressions for Riemannian tensors from eqns \eqref{Rrtrt}- \eqref{Rphrtph} into
the eqn \eqref{Karmarkar} we have
\begin{eqnarray}
0&=&b^{2}{\dot{s}}^2b^{3}\left(\frac{a^{\prime\prime}}{a}-2\frac{a^{\prime}}{a}\frac{b\prime}{b}+\frac{a^{\prime\,^2}}{a^{2}}-\frac{a^{\prime}}{ra}\right)
-r^{2}b^{3}s\ddot{s}\left(\frac{b^{\prime\prime}}{b}-2\frac{b^{\prime\,^{2}}}{b^{2}} -\frac{b^{\prime}}{rb}\right)\nonumber\\
&+&r^{2}aa^{\prime}b^{\prime\prime}\left(\frac{b^{\prime}}{b}+\frac{1}{r} \right)-r^{2}aba^{\prime\prime}\left(\frac{b^{\prime\,^{2}}}{b^2}+2\frac{b^{\prime}}{rb}
\right)+raba^{\prime}\left(\frac{b^{\prime}}{rb}+2\frac{b^{\prime\,^{2}}}{b^2}\right). \label{Karmarkar1}
\end{eqnarray}
For a given form of metric function $b(r)$ \eqref{br}, the class {\bf{I}} condition in equation \eqref{Karmarkar1} is nonlinear.
A physical relevant collapsing model must satisfy \eqref{matching} 
and \eqref{Karmarkar1} simultaneously.
It must be noted that simplest choice of solutions of \eqref{matching} is a linear solution \,\cite{Banerjee}
\begin{eqnarray}
s(t)&=&-C_{Z} \,t , \label{s(t)}
\end{eqnarray}
 $C_Z>0$. 
 The form of the other metric function $a(r)$ is obtained by using equation \eqref{br} and \eqref{s(t)} into the class $1$ condition \eqref{Karmarkar1}
\begin{eqnarray}
a(r)&=&\frac{1}{2\sqrt{2 C_{3}C_{4}}}[C_4^{2}\left(C_1b(r)+4C_2C_3\right)^2-4C_Z^2]^{1/2} \label{a(r)}
\end{eqnarray}
where $C_1$ and $C_2$ are integration constants. 
Surprisingly, the quantity in the numerator inside the square root, arises naturally from the 
matching of the Ricci scalar (and it's derivative), given in \eqref{Ricci_junction}
These forms of the solutions of the gravitational potentials are same as obtained in\,\cite{SCJ} for 
shear free spacetime. In \cite{SCJ}, it has been shown that for the static case, Karmarkar condition 
together with the pressure isotropy yields the Schwarzschild\,\cite{KS} like form of the metric functions.
Also, it has been shown that these set of gravitational potentials are the special class of those found in\,\cite{ON}.
Thus, although we have assume this particular form of $\Delta$ \eqref{Delta1} for the mathematical simplicity, 
represents the physically viable solutions.

\begin{figure}[t!h]
	\begin{subfigure}{0.333\textwidth}
		\centering
		\includegraphics[width=\linewidth]{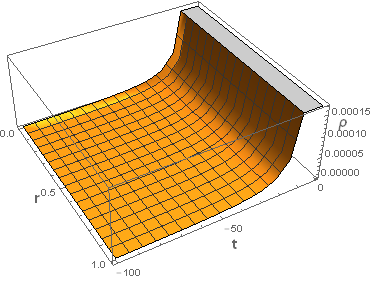}
		\caption{}
		\label{fig:rho11}
	\end{subfigure}
	\begin{subfigure}{0.333\textwidth}
		\centering
		\includegraphics[width=\linewidth]{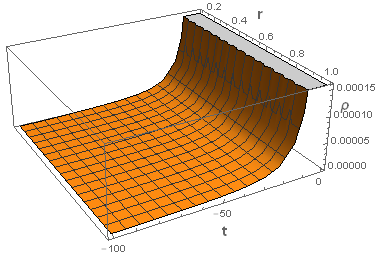}
		\caption{}
		\label{fig:rho12}
	\end{subfigure}
	\begin{subfigure}{0.333\textwidth}
		\centering
		\includegraphics[width=\linewidth]{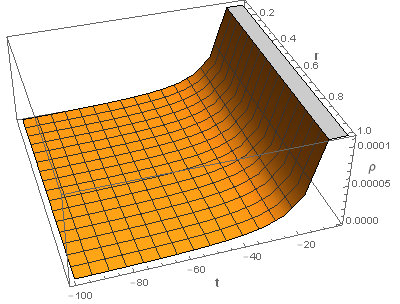}
		\caption{}
		\label{fig:rho13}
	\end{subfigure}
	\caption{(a), (b) \& (c) shows the plots of the density $\rho$ \eqref{rhoabf}
		w.r.t. time $t$ and radial $r$ coordinates for $f(R)=R+\lambda R^{2}$ with $\lambda=5$, $f(R)=R^{1+\epsilon}$ with $\epsilon=0.01$ and $f(R)=R+\lambda \left[\exp\left(-\gamma R\right)-1\right]$ with $\lambda=0.1$ \& $\gamma=0.0002$ respectively. For each of these $f(R)$ models, it remains regular as well as positive throughout the collapse.}
\end{figure}

\begin{figure}[h!]
	\begin{subfigure}{0.333\textwidth}
		\centering
		\includegraphics[width=\linewidth]{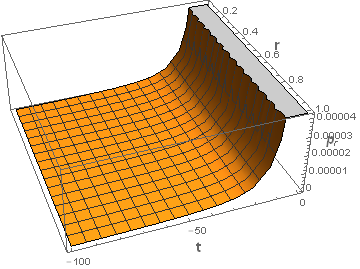}
		\caption{}
		\label{fig:pr11}
	\end{subfigure}
	\begin{subfigure}{0.333\textwidth}
		\centering
		\includegraphics[width=\linewidth]{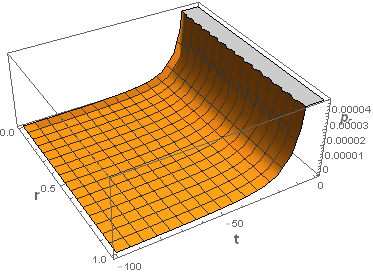}
		\caption{}
	\label{fig:pr12}
	\end{subfigure}
	\begin{subfigure}{0.333\textwidth}
		\centering
		\includegraphics[width=\linewidth]{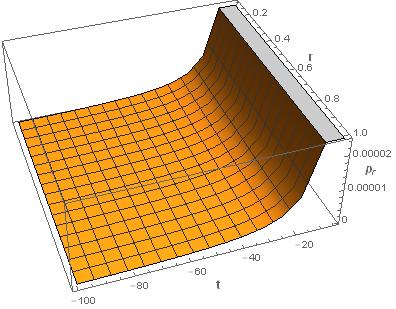}
		\caption{}
		\label{fig:pr13}
	\end{subfigure}
	\caption{(a), (b) \& (c) shows the plots of the radial pressure $p_{r}$ \eqref{prabf}
		w.r.t. time $t$ and radial $r$ coordinates for $f(R)=R+\lambda R^{2}$ with $\lambda=5$, $f(R)=R^{1+\epsilon}$ with $\epsilon=0.01$ and $f(R)=R+\lambda \left[\exp\left(-\gamma R\right)-1\right]$ with $\lambda=0.1$ \& $\gamma=0.0002$ respectively. For each of these $f(R)$ models, it remains regular as well as positive throughout the collapse.}
\end{figure}

\begin{figure}[h!]
	\begin{subfigure}{0.333\textwidth}
		\centering
	\includegraphics[width=\linewidth]{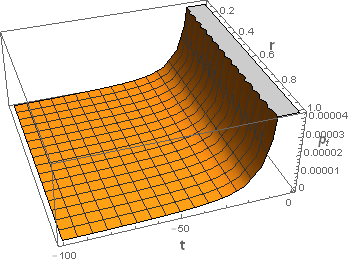}
		\caption{}
		\label{fig:pt11}
	\end{subfigure}
	\begin{subfigure}{0.333\textwidth}
		\centering
	\includegraphics[width=\linewidth]{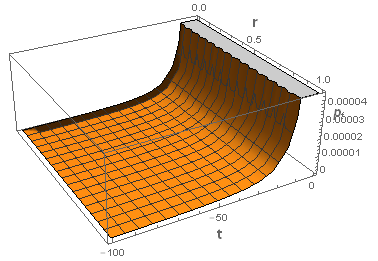}
		\caption{}
		\label{fig:pt12}
	\end{subfigure}
	\begin{subfigure}{0.333\textwidth}
		\centering
		\includegraphics[width=\linewidth]{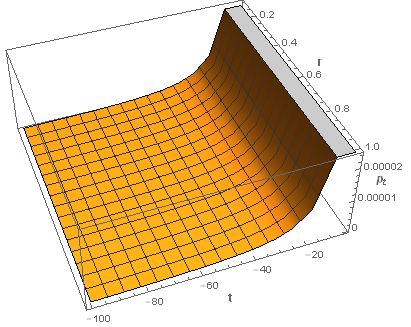}
		\caption{}
		\label{fig:pt13}
	\end{subfigure}
	\caption{(a), (b) \& (c) shows the plots of the tangential pressure $p_{t}$ \eqref{ptabf}
		w.r.t. time $t$ and radial $r$ coordinates for $f(R)=R+\lambda R^{2}$ with $\lambda=5$, $f(R)=R^{1+\epsilon}$ with $\epsilon=0.01$ and $f(R)=R+\lambda \left[\exp\left(-\gamma R\right)-1\right]$ with $\lambda=0.1$ \& $\gamma=0.0002$ respectively. For each of these $f(R)$ models, it remains regular as well as positive throughout the collapse.}
\end{figure}

\begin{figure}[h!]
	\begin{subfigure}{0.333\textwidth}
		\centering
		\includegraphics[width=\linewidth]{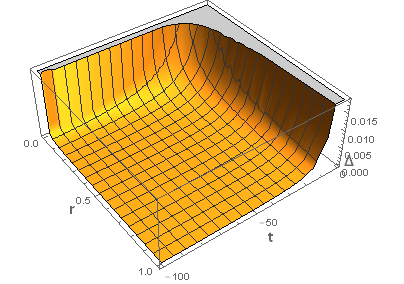}
		\caption{}
	\label{fig:delta11}
	\end{subfigure}
\begin{subfigure}{0.333\textwidth}
		\centering
		\includegraphics[width=\linewidth]{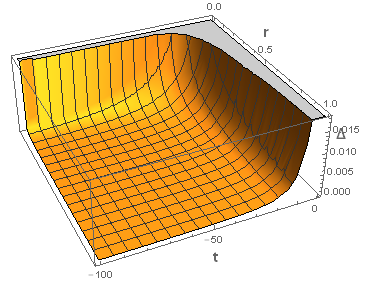}
		\caption{}
		\label{fig:delta12}
	\end{subfigure}
	\begin{subfigure}{0.333\textwidth}
		\centering
	\includegraphics[width=\linewidth]{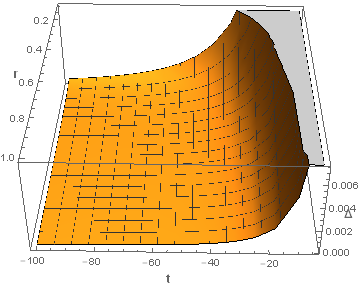}
		\caption{}
		\label{fig:delta13}
	\end{subfigure}
\caption{(a), (b) \& (c) shows the plots of the pressure anisotropy $\Delta$ \eqref{Delta1}
		w.r.t. time $t$ and radial $r$ coordinates for $f(R)=R+\lambda R^{2}$ with $\lambda=5$, $f(R)=R^{1+\epsilon}$ with $\epsilon=0.01$ and $f(R)=R+\lambda \left[\exp\left(-\gamma R\right)-1\right]$ with $\lambda=0.1$ \& $\gamma=0.0002$ respectively. For each of these $f(R)$ models, it remains regular as well as positive throughout the collapse.}
\end{figure}

\begin{figure}[h!]
	\begin{subfigure}{0.333\textwidth}
		\centering
		\includegraphics[width=\linewidth]{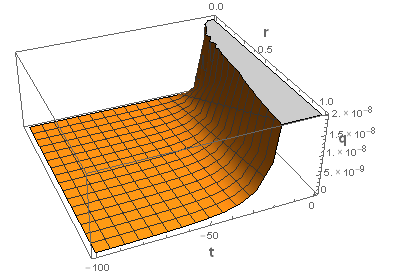}
		\caption{}
		\label{fig:q11}
\end{subfigure}
	\begin{subfigure}{0.333\textwidth}
		\centering
		\includegraphics[width=\linewidth]{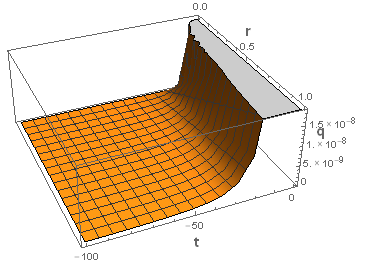}
		\caption{}
		\label{fig:q12}
	\end{subfigure}
	\begin{subfigure}{0.333\textwidth}
		\centering
		\includegraphics[width=\linewidth]{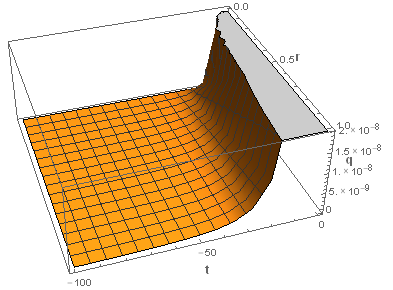}
		\caption{}
		\label{fig:q13}
	\end{subfigure}
	\caption{(a), (b) \& (c) shows the plots of the heat flux $q$ \eqref{qabf}
		w.r.t. time $t$ and radial $r$ coordinates for $f(R)=R+\lambda R^{2}$ with $\lambda=5$, $f(R)=R^{1+\epsilon}$ with $\epsilon=0.01$ and $f(R)=R+\lambda \left[\exp\left(-\gamma R\right)-1\right]$ with $\lambda=0.1$ \& $\gamma=0.0002$ respectively. For each of these $f(R)$ models, it remains positive throughout the collapse.}
\end{figure}

\begin{figure}[h!]
\begin{subfigure}{0.333\textwidth}
\centering
\includegraphics[width=\linewidth]{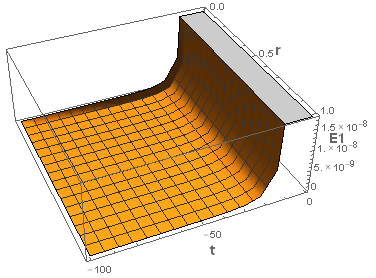}
\caption{}
\label{fig:E11}
\end{subfigure}
\begin{subfigure}{0.333\textwidth}
\centering
\includegraphics[width=\linewidth]{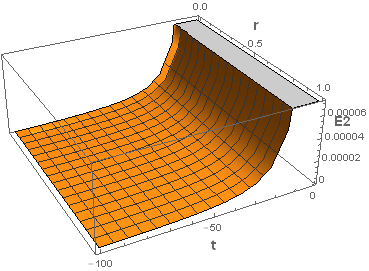}
\caption{}
\label{fig:E12}
\end{subfigure}
\begin{subfigure}{0.333\textwidth}
\centering
\includegraphics[width=\linewidth]{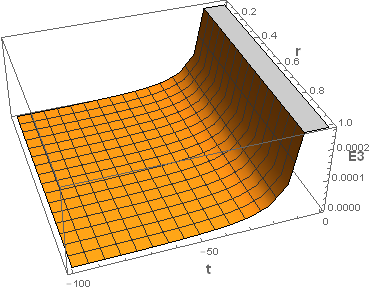}
\caption{}
\label{fig:E13}
\end{subfigure}
\caption{(a), (b) and (c) shows the plots of energy conditions $E1$, $E2$ and $E3$ w.r.t. time $t$ and radial $r$ coordinates for $f(R)=R+\lambda R^{2}$ with $\lambda=5$ respectively.}
\end{figure}

It is now instructive to rewrite the physical quantities of the matter cloud
in terms of the metric variables for a better understanding of the dynamics of spacetime 
during the collapse process. These expressions have been written 
in detail in the Appendix.
The boundary condition \eqref{matching} in the view of these equations
in the Appendix, \eqref{prabf}-\eqref{qabf} becomes
\begin{eqnarray}
2s\,\ddot{s}+\dot{s}^2-2x\dot{s}+b\,s\,T^{D}_{rt}&=&y-T^{D}_{rr} ,\label{match}
\end{eqnarray}
where $T^{D}_{rr}$ and $T^{D}_{rt}$ are 
given by equations \eqref{darkT_rr} and \eqref{darkT_tr} respectively and the quantities $x$ and $y$
are
\begin{eqnarray}
x=\left(\frac{a^{\prime}}{b} \right)_{\Sigma},\label{x} ~~~~
y=\left(\frac{a^2}{b^2}\left[\frac{b^{\prime\,^{2}}}{b^2}+\frac{2}{r}\frac{b^{\prime}}{b}+\frac{2a^{\prime}b^{\prime}}{ab}+\frac{2}{r}\frac{a^{\prime}}{a}\right] \right)_{\Sigma}. \label{y}
\end{eqnarray}
The metric functions $a(r)$ and $b(r)$ should not vanish during the collapsing phenomena, since otherwise the metric shall become degenerate.
This also implies that their signatures remain unchanged.  
For second metric potential to be positive i.e. $a(r)>0$ 
we must have, from \eqref{a(r)}, that
\begin{eqnarray}
C_{Z}^2<C_{4}^{2}\left[\frac{C_{1}}{C_{3}r^{2}+2C_{4}}-2C_{2}C_{3}\right]^2.
\end{eqnarray}
This equation also implies that 
at the center of the cloud, $r=0$, we must have $C_{Z}<C_{1}-2C_{2}C_{3}C_{4}$. 
\\
\\
The graphical representations of the physical quantities \eqref{rhoabf}-\eqref{qabf} shows that they are
well defined throughout the stellar collapse for both the $f(R)$ models.
Figs. \ref{fig:rho11}\,, \ref{fig:rho12} \& \ref{fig:rho13}\,, \ref{fig:pr11}\,,
\ref{fig:pr12} \& \ref{fig:pr13}\,, \ref{fig:pt11}\,, \ref{fig:pt12} \& \ref{fig:pt13}\,, \ref{fig:delta11}\,, \ref{fig:delta12} \& \ref{fig:delta13}
shows that the density, radial pressure, tangential pressure and pressure anisotropy
are positive and regular throughout the collapse for all the $f(R)$ models and the parameters
considered here. Also, as seen 
from the Figs. \ref{fig:q11}\,, \ref{fig:q12} \& \ref{fig:q13}\,,
the heat flux increase as the collapse starts and remains positive throughout the collapse
for these cases.
\begin{figure}[h!]
\begin{subfigure}{.5\textwidth}
\centering
\includegraphics[width=\linewidth]{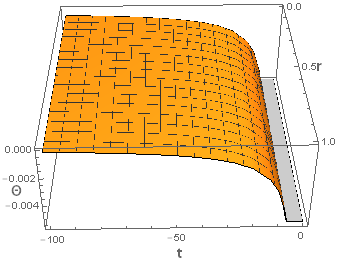}
\caption{}
\label{fig:Theta}
\end{subfigure}\begin{subfigure}{0.5\textwidth}
\centering
\includegraphics[width=\linewidth]{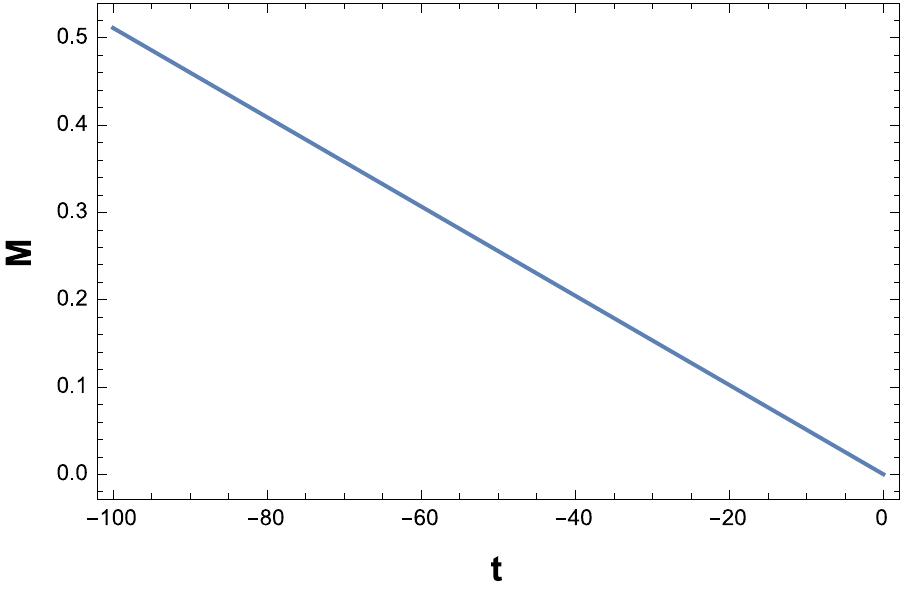}
\caption{}
\label{fig:m}
\end{subfigure}
\caption{ (a) Plot of the expansion scalar $\Theta$ \eqref{Thetaabf}
w.r.t. time $t$ and radial $r$ coordinates. 
At the beginning of collapse $\Theta$ has zero value and it starts decreasing and remains negative throughout the collapse. (b) Plot of the mass of the collapsing star \eqref{mabf} w.r.t. time $t$, and it shows that mass radiates linearly.}
\end{figure}
\clearpage
A related quantity of importance in this study is
the  total luminosity 
visible to an observer at infinity, which
may be defined in terms of the mass loss from the boundary surface:
\begin{eqnarray}
L_{\infty}&=& -\left( \frac{dm}{dv}\right)_{\Sigma}=\left[\frac{r^2\,s^{2}\,b^{2}\,p_{r}}{2}\left(1+\frac{r\,b^{\prime}}{b}+\frac{r\,b\,\dot{s}}{a}\right)^2\right]_{\Sigma},\label{LInfinity}
\end{eqnarray}
where we have used the equations \eqref{pr}, \eqref{mass} and \eqref{Ktt+-}.
Now, as soon as the black hole is formed, by definition, the luminosity 
of the surface is zero. From the above equation, 
this implies that sufficient condition for the formation of a black hole
is 
\begin{eqnarray}
\left[1+\frac{r\,b^{\prime}}{b}+\frac{r\,b\,\dot{s}}{a}\right]_{\Sigma}=0.
\end{eqnarray}
Naturally,
for any static observer at asymptotic infinity, 
the redshift diverges at the time of formation of the black hole.
\begin{figure}[h!]
	\begin{subfigure}{0.333\textwidth}
		\centering
		\includegraphics[width=\linewidth]{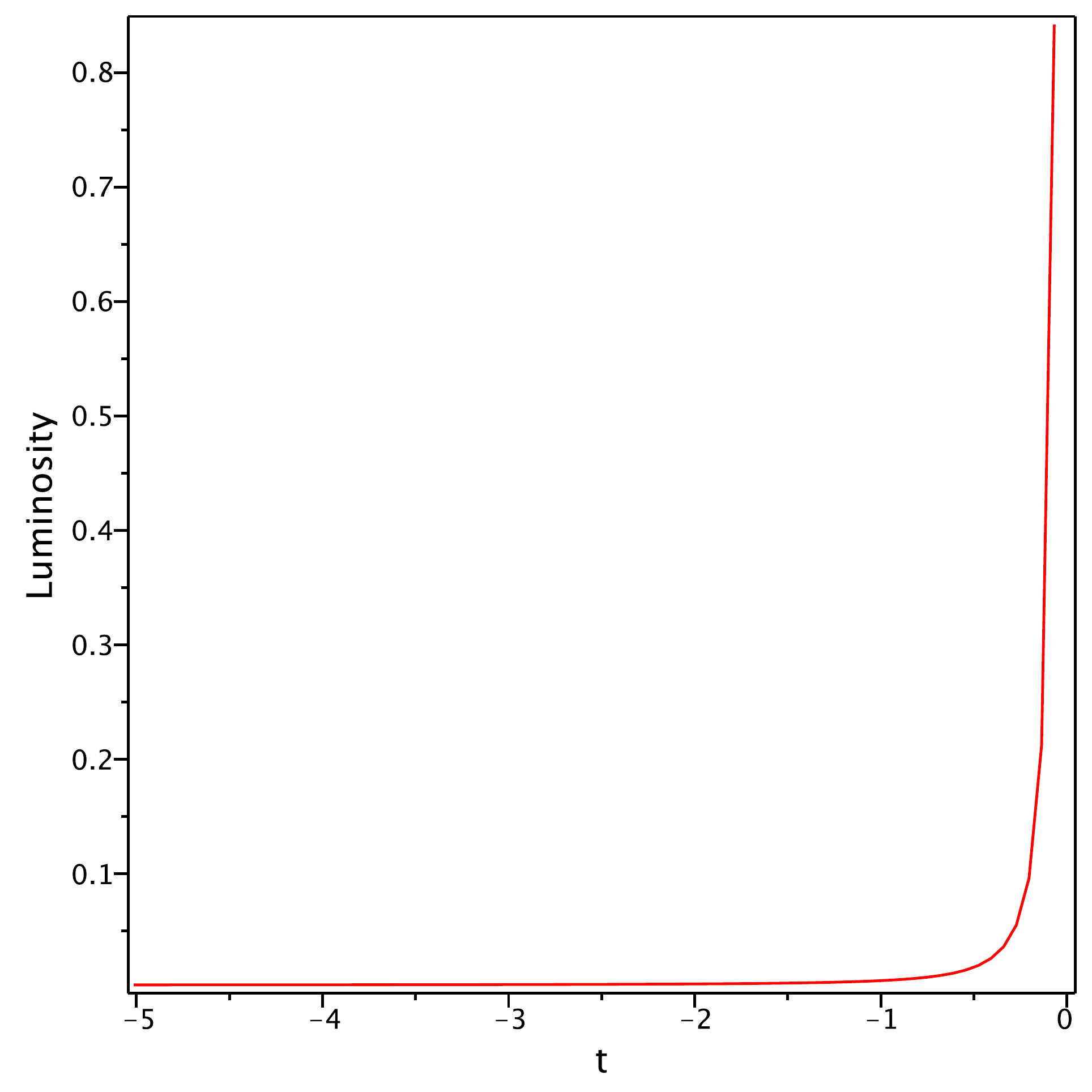}
		\caption{}
		\label{fig:Lumf1}
	\end{subfigure}
	\begin{subfigure}{0.333\textwidth}
		\centering
	\includegraphics[width=\linewidth]{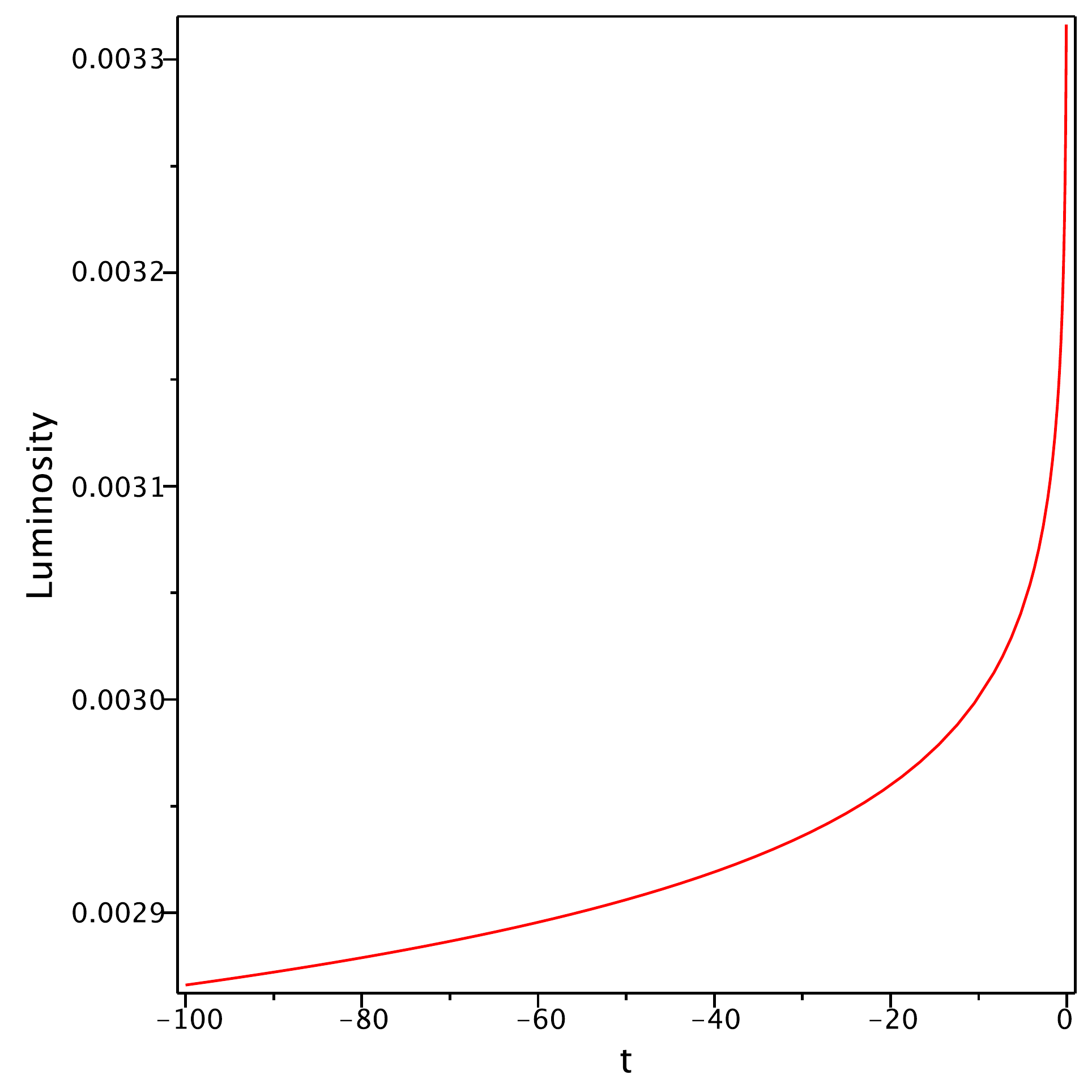}
		\caption{}
		\label{fig:Lumf2}
	\end{subfigure}
	\begin{subfigure}{0.333\textwidth}
		\centering
		\includegraphics[width=\linewidth]{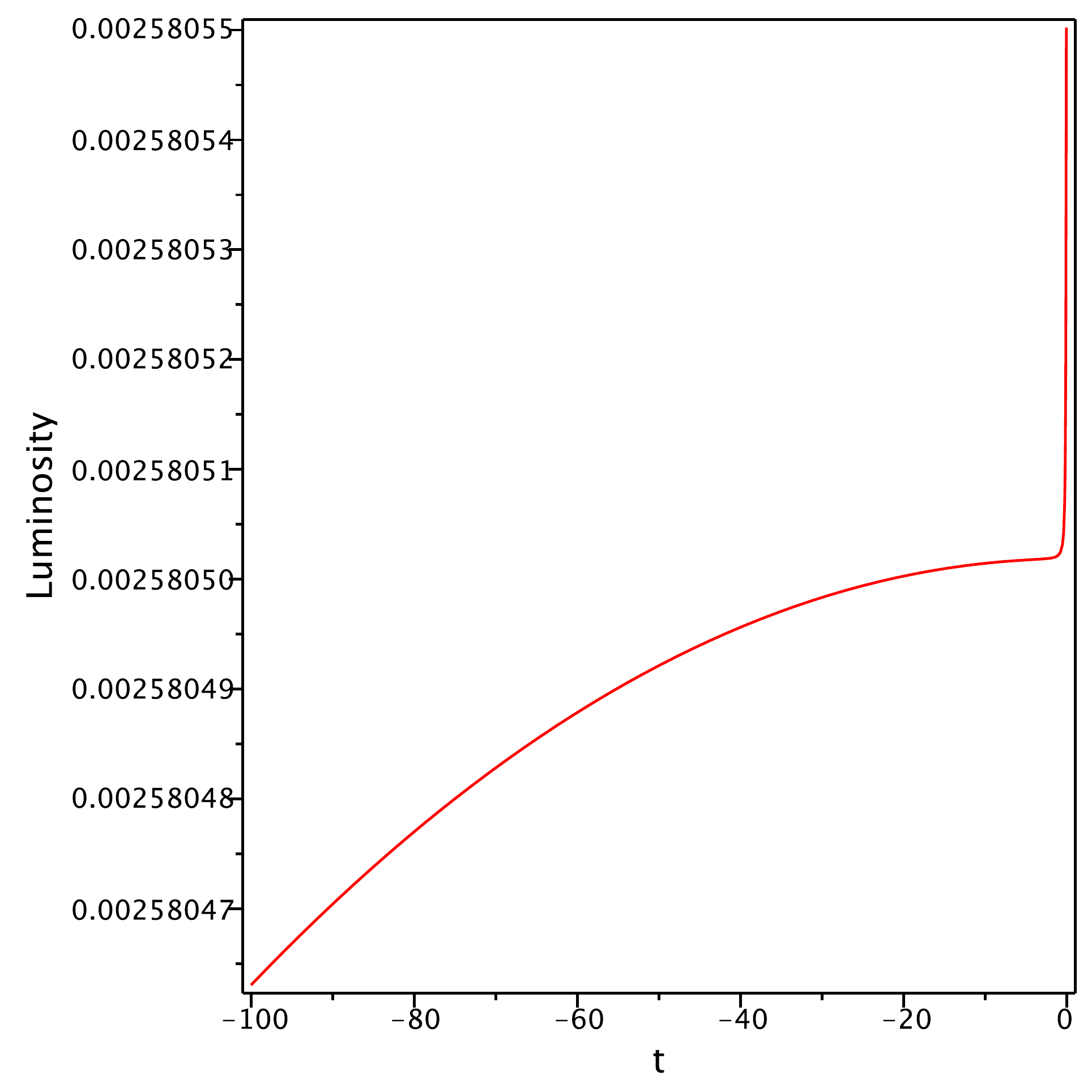}
		\caption{}
		\label{fig:Lumf3}
	\end{subfigure}
	\caption{(a), (b) \& (c) shows the plots of the luminosity \eqref{LInfinity} at $r=r_{\Sigma}=1$, w.r.t. time $t$ for $f(R)=R+\lambda R^{2}$ with $\lambda=5$, $f(R)=R^{1+\epsilon}$ with $\epsilon=0.01$ and $f(R)=R+\lambda \left[\exp\left(-\gamma R\right)-1\right]$ with $\lambda=0.1$ \& $\gamma=0.0002$ respectively.}
\end{figure}
\\
To show that these spacetime solutions are physically viable,
we show that they satisfy the energy conditions as well.
Indeed, all the energy conditions
namely weak (W), null (N), 
dominant (D) and strong (S) hold good for the collapsing star.
In the following we list these conditions
\cite{KS,Chan1997}\\
{\bf{E1\,:}}\, $\left(\rho+p_{r}\right)^2-4q^{2}\,\geq \,0$\hspace{3.2cm}(D/S/W)\\
{\bf{E2\,:}}\, $\rho-p_{r}\,\geq\,0\,\,\,\, $\hspace{4.3cm} (D)\\
{\bf{E3\,:}}\, $\rho-p_{r}-2p_{t}+\sqrt{\left(\rho+p_{r}\right)^2-4q^{2}}\,\geq\,0 $\hspace{.2cm} (D)\\
{\bf{E4\,:}}\, $\rho-p_{r}+\sqrt{\left(\rho+p_{r}\right)^2-4q^{2}}\,\geq\,0\,\,\,\, $\hspace{.9cm} (W/D)\\
{\bf{E5\,:}}\, $\rho-p_{r}+2p_{t}+\sqrt{\left(\rho+p_{r}\right)^2-4q^{2}}\,\geq\,0$\hspace{.2cm} (D/W/S)\\
{\bf{E6\,:}}\, $2p_{t}+\sqrt{\left(\rho+p_{r}\right)^2-4q^{2}}\,\geq\,0$\hspace{1.7cm} (S)\\
The star should also satisfy\\
{\bf{E7\,:}}\, $\rho>0$,\, $p_{r}>0$,\, $p_{t}>0$, and $\rho^{\prime}<0$,\, $p_r^{\prime}<0$,\, $p_{t}^{\prime}<0$.\\
It is clear from the above conditions that 
{\bf{E1}},\, {\bf{E2}},\,{\bf{E3}} \& {\bf{E7}} are enough to validate the physical conditions existing inside the star. 
For the radiating- collapsing stellar models in $f(R)$ gravity,
Figs. \ref{fig:E11}, \ref{fig:E12} \& \ref{fig:E13} show 
that the energy conditions are positive and regular throughout 
the interior of the star.
\begin{figure}[h!]
\begin{subfigure}{0.333\textwidth}
\centering
\includegraphics[width=\linewidth]{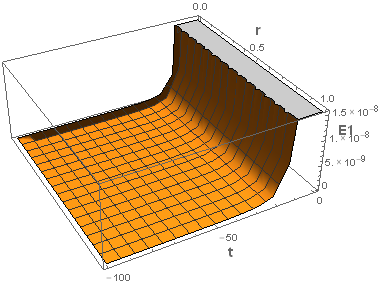}
\caption{}
\label{fig:E21}
\end{subfigure}
\begin{subfigure}{0.333\textwidth}
\centering
\includegraphics[width=\linewidth]{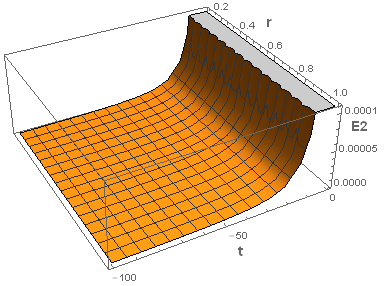}
\caption{}
\label{fig:E22}
\end{subfigure}
\begin{subfigure}{0.333\textwidth}
\centering
\includegraphics[width=\linewidth]{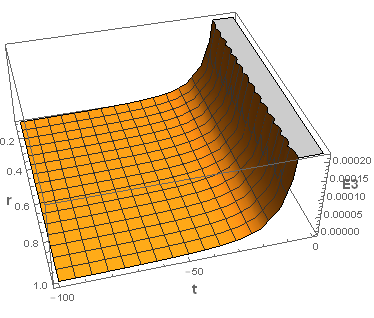}
\caption{}
\label{fig:E23}
\end{subfigure}
\caption{(a), (b) and (c) shows the plots of energy conditions $E1$, $E2$ and $E3$ w.r.t. time $t$ and radial $r$ coordinates for $f(R)=R^{1+\epsilon}$ with $\epsilon=0.01$ respectively.}
\end{figure}
\\
\begin{figure}[t!]
	\begin{subfigure}{0.333\textwidth}
		\centering
		\includegraphics[width=\linewidth]{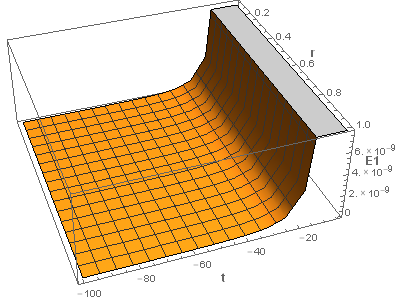}
		\caption{}
	\label{fig:E31}
	\end{subfigure}
	\begin{subfigure}{0.333\textwidth}
		\centering
		\includegraphics[width=\linewidth]{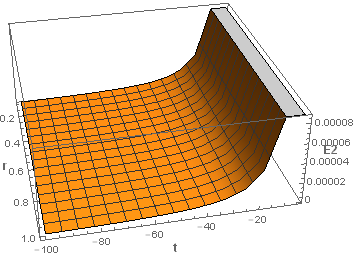}
		\caption{}
		\label{fig:E32}
	\end{subfigure}
	\begin{subfigure}{0.333\textwidth}
		\centering
		\includegraphics[width=\linewidth]{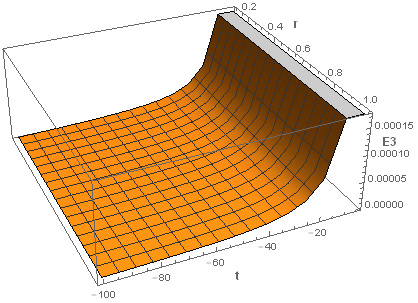}
		\caption{}
		\label{fig:E33}
	\end{subfigure}
	\caption{(a), (b) \& (c) shows the plots of energy conditions $E1$, $E2$ and $E3$ w.r.t. time $t$ and radial $r$ coordinates for $f(R)=R+\lambda \left[\exp\left(-\gamma R\right)-1\right]$ with $\lambda=0.1$ \& $\gamma=0.0002$ respectively.}
\end{figure}
\subsection{Stability Criteria}\label{sec3}

The study of dynamical instability\,(stability) of spherical stellar system shows 
that for adiabatic index $\Gamma<4/3$ $\left(\Gamma>4/3\right)$ the stellar system becomes 
unstable (stable) as the weight of the stellar system increase much faster 
(remains less than) than that of its pressure\,\cite{SChandra}.
Also, the causality condition imposes certain constraints on 
the dynamics of the stellar system such that inside the star,
the radial $V_{r}$ and the transverse $V_{t}$ components of the
speed of sound should be less than
the speed of the light ($c=1$), so that $0\leq V_{r}\leq 1$ and $0\leq V_{t}\leq 1$\,\cite{LH1992}.
Thus, to check the stability/instability of the collapsing stellar system,
we need to study the behavior of the important physical quantities, 
adiabatic index, sound of the speed which are defined as \cite{Santos,LH1992,GP_RC}
\begin{eqnarray}
\Gamma_{eff}&=&\left[\frac{\partial ( \ln p_{r})}{\partial ( \ln \rho)}\right]_{\Sigma}
=\left[\left(\frac{\dot{p}_{r}}{p_{r}}\right) \left(\frac{\dot{\rho}}{\rho}\right)\right]_{\Sigma}\label{adiabatic}\\
Vr&=&\frac{d\,p_{r}}{d\rho}\,\, ,\hspace{1cm}Vt=\frac{d\,p_{t}}{d\rho}\label{speed}
\end{eqnarray}
\begin{figure}[h!]
	\begin{subfigure}{0.333\textwidth}
	\centering
		\includegraphics[width=\linewidth]{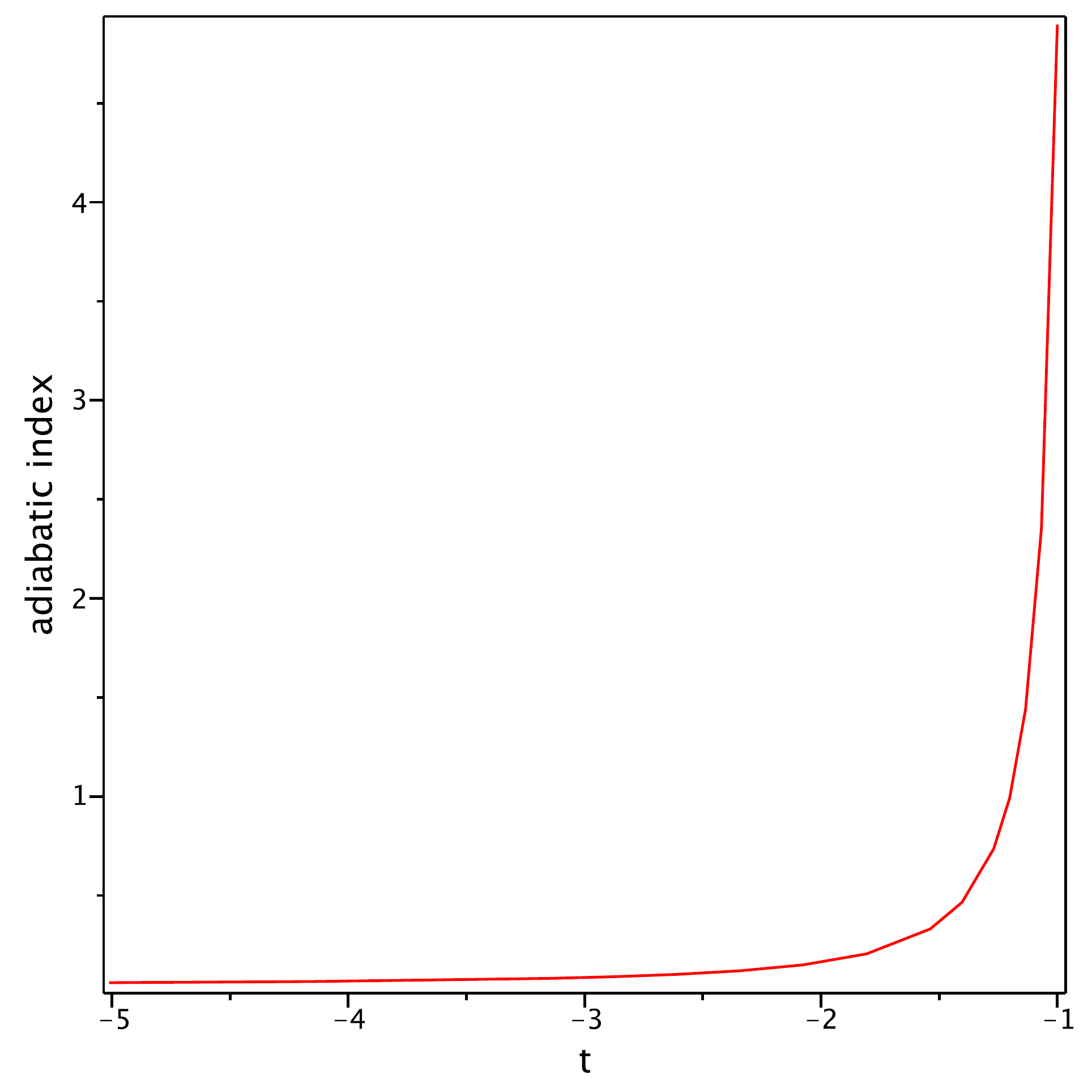}
		\caption{}
		\label{fig:adf1}
	\end{subfigure}
\begin{subfigure}{0.333\textwidth}
		\centering
		\includegraphics[width=\linewidth]{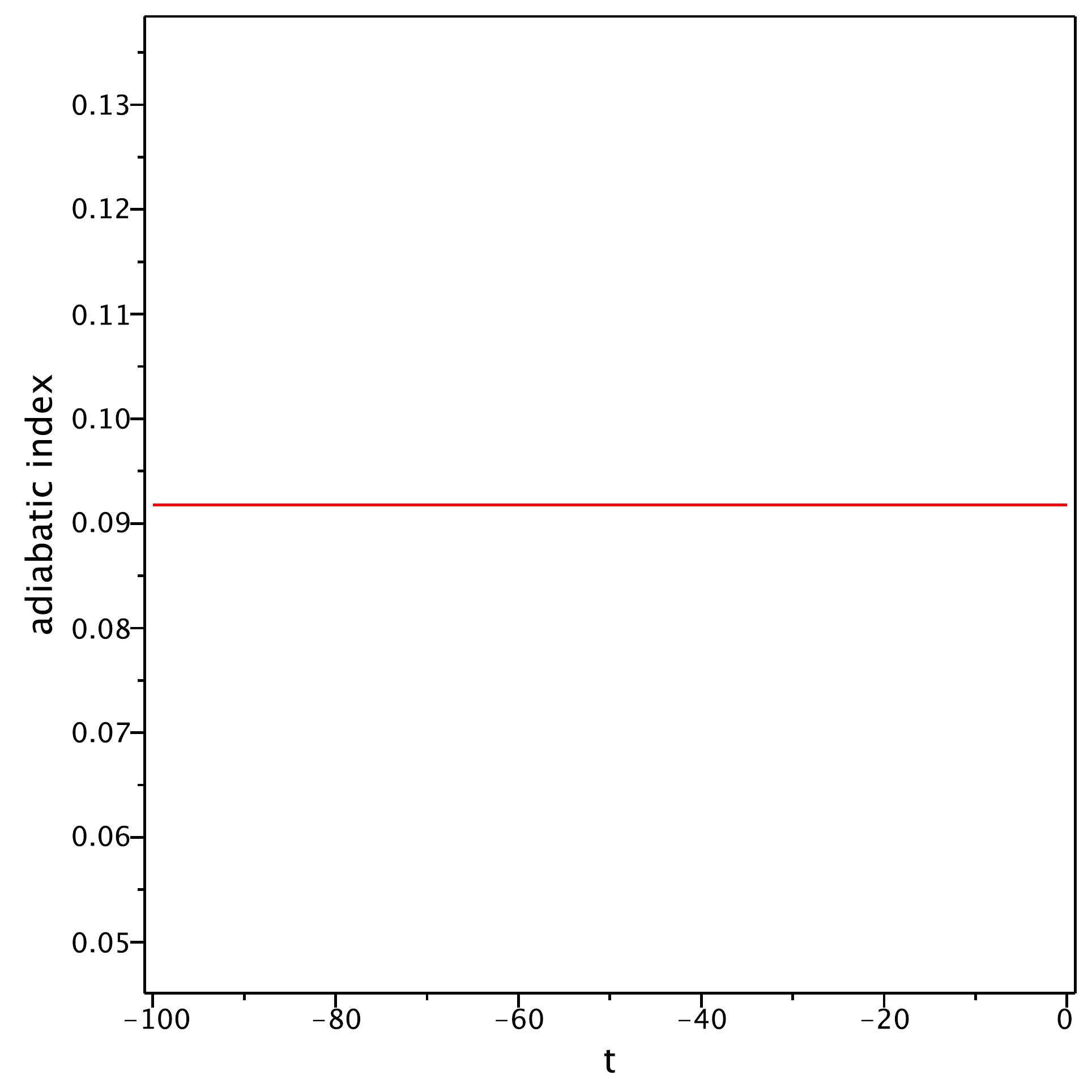}
		\caption{}
		\label{fig:adf2}
	\end{subfigure}
	\begin{subfigure}{0.333\textwidth}
		\centering
		\includegraphics[width=\linewidth]{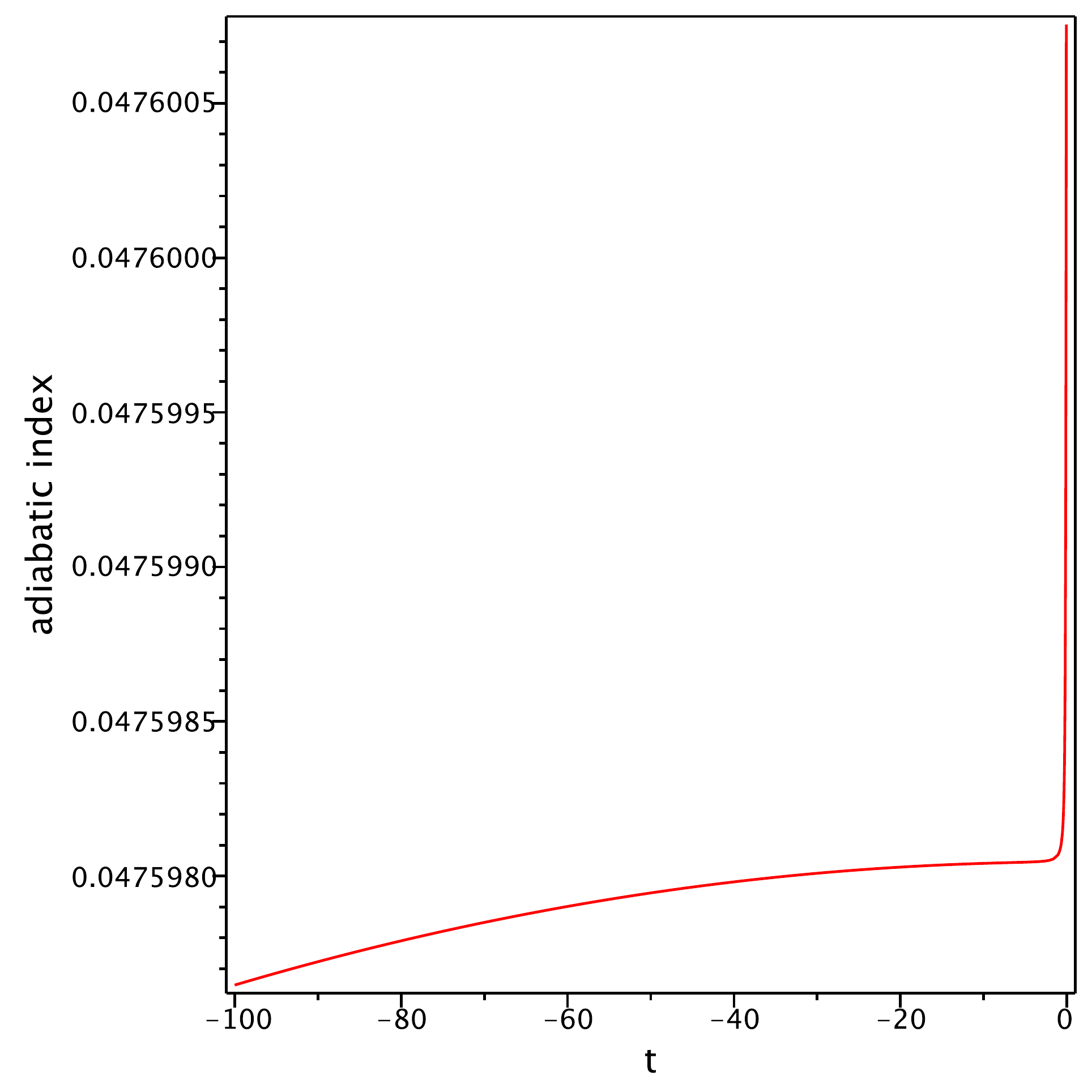}
		\caption{}
		\label{fig:adf3}
	\end{subfigure}
	\caption{(a), (b) \& (c) shows the plots of adiabatic index at $r=r_{\Sigma}=1$, w.r.t. time $t$ for $f(R)=R+\lambda R^{2}$ with $\lambda=5$, $f(R)=R^{1+\epsilon}$ with $\epsilon=0.01$ and $f(R)=R+\lambda \left[\exp\left(-\gamma R\right)-1\right]$ with $\lambda=0.1$ \& $\gamma=0.0002$ respectively}
\end{figure}
Although stability may be understood from the behavior of
the pressure and density variables, 
the quantities in \eqref{adiabatic} and \eqref{speed} 
are considered to be better to establish
stability. 
For $f(R)=R+\lambda R^{2}$ model with $\lambda=5$, Figs. \ref{fig:Lumf1} and \ref{fig:adf1}
shows that the total luminosity and the adiabatic index are positive and increasing.
Note that the adiabatic index attains a maximum value where the luminosity is maximum. 
This behavior of the luminosity and 
adiabatic index can be interpreted as follows. 
Any static observer at asymptotic infinity will see an exponentially radiating radial source until a time when luminosity reaches its maximum value after which it instantaneously turn off.
 This is due to the fact that 
 the total mass of the star radiates linearly as seen from the Fig. \ref{fig:m}
  and when the star reaches its maximum luminosity, there is no mass left to radiates 
  and hence the observer at rest at infinity will see sudden turn off of the light source.
  The similar kind of behavior were obtained in\,\cite{GP_RC}. Fig. \ref{fig:adf1} 
  shows that the effective adiabatic index is positive and less than $4/3$ which implies that the considered stellar system
is unstable and representing the collapsing scenario\,\cite{SChandra}. 
For $f(R)=R^{1+\epsilon}$, and $f(R)=R+\lambda\,\left[
\exp{(-\sigma R)}-1\right]$ models, similar behavior of luminosity is obtained as 
that of for the first model.
Fig. \ref{fig:adf2} shows that the effective adiabatic 
index is constant function of time, and is positive and less than $4/3$,
 which implies it represents the collapsing scenario. 
As we have shown graphically that the star radiates all 
its mass before reaching at the singularity. So, there are no trapped surfaces formed during the collapse.
Which implies that neither the black hole nor naked singularity are the end state of the collapse. 
\subsection{Thermal Properties }\label{sec5}
Earlier studies have shown that relaxation 
effects are important to understand dissipative gravitational collapse 
\cite{Martinez1996, Herrera1996, Herrera1997, GMM1999, Herrera2004}.
To study the temperature profiles, 
we consider the transport equation for the metric 
\eqref{1eq1} given by\,\cite{Maartens1995, Martinez1996, Israel1979}
\begin{eqnarray}
\tau h_{\mu}^{\nu}\dot{q}_{\nu}+q_{\mu}&=&-k\left( h_{\mu}^{\nu} \nabla_{\nu} T+T\dot{u}_{\mu} \right)\label{tempgen}\\
\tau \left(qbs\right)_{,t}+q\,a\,b\,s&=&-\frac{k\left(aT \right)_{,r}}{bs}, \label{tempabf}
\end{eqnarray}
where, $\alpha>0$,\,$\beta>0$,\,
$\gamma>0$ and $\sigma>0$, \& $h^{\mu\nu}=g^{\mu\nu}+u^{\mu}u^{\nu}$.
Also,
\begin{eqnarray}
 \tau_{c}=\left({\alpha/\gamma}\right)\,(T)^{-\sigma}\hspace{0.2cm},\hspace{0.6cm} 
 k=\gamma\,\tau_{c}\, T^{3}\hspace{0.2cm},\hspace{0.6cm}  \tau=\tau_{c}\,(\beta\,\gamma)/\alpha \label{tkt}
\end{eqnarray}
where  $\tau_{c}$ is the mean collision time, $k$ is thermal conductivity and $\tau$ represents the relaxation time respectively \cite{Martinez1996}\cite{GG2001}. 
The quantity $\tau$ measures the strength of relaxational effects and is called the causality index.
The values $\tau=0$ or $\beta=0$ represents the noncausal temperature profile.
\begin{figure}[h!]
	\begin{subfigure}{0.333\textwidth}
		\centering
		\includegraphics[width=\linewidth]{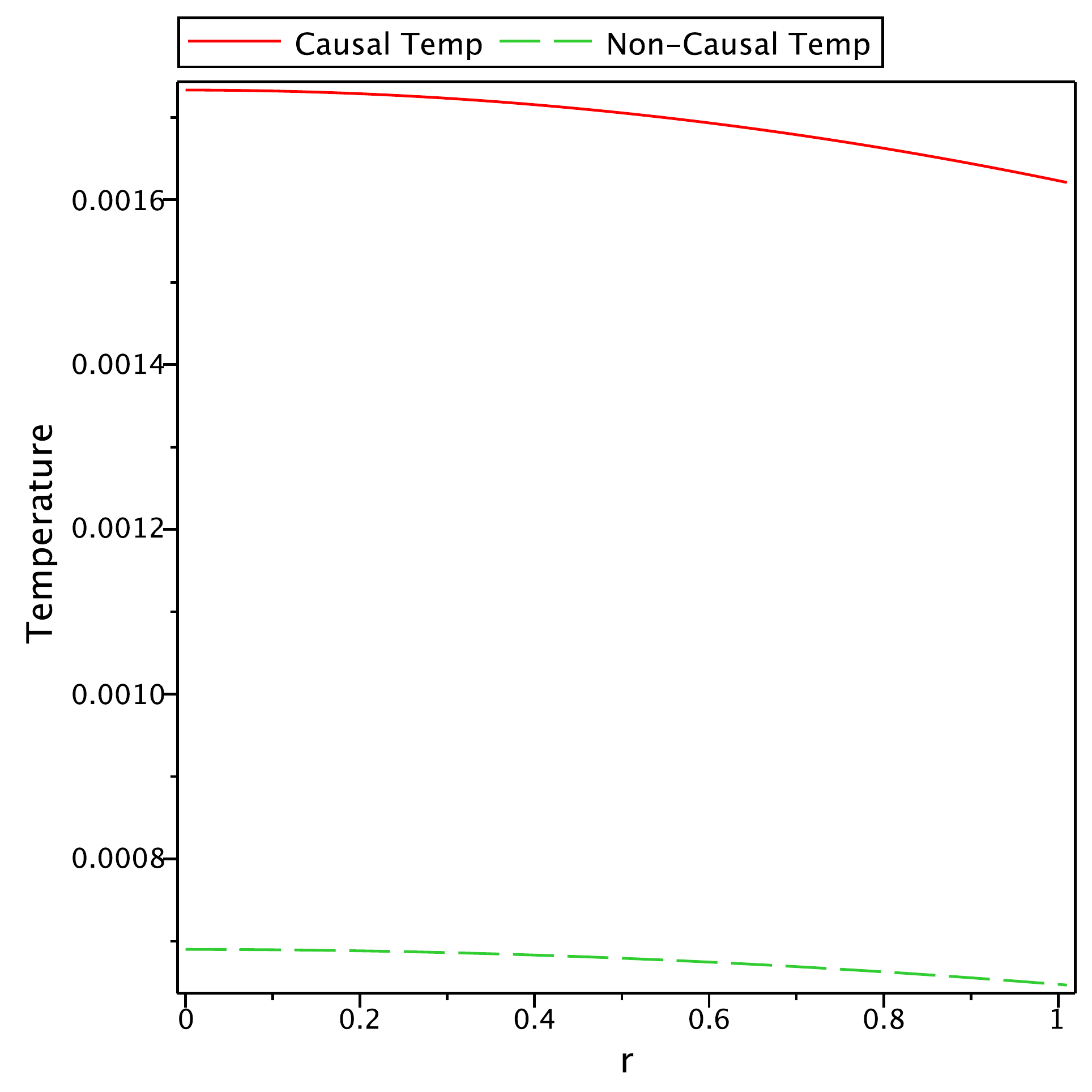}
		\caption{}
		\label{fig:Temp}
	\end{subfigure}
	\begin{subfigure}{0.333\textwidth}
		\centering
		\includegraphics[width=\linewidth]{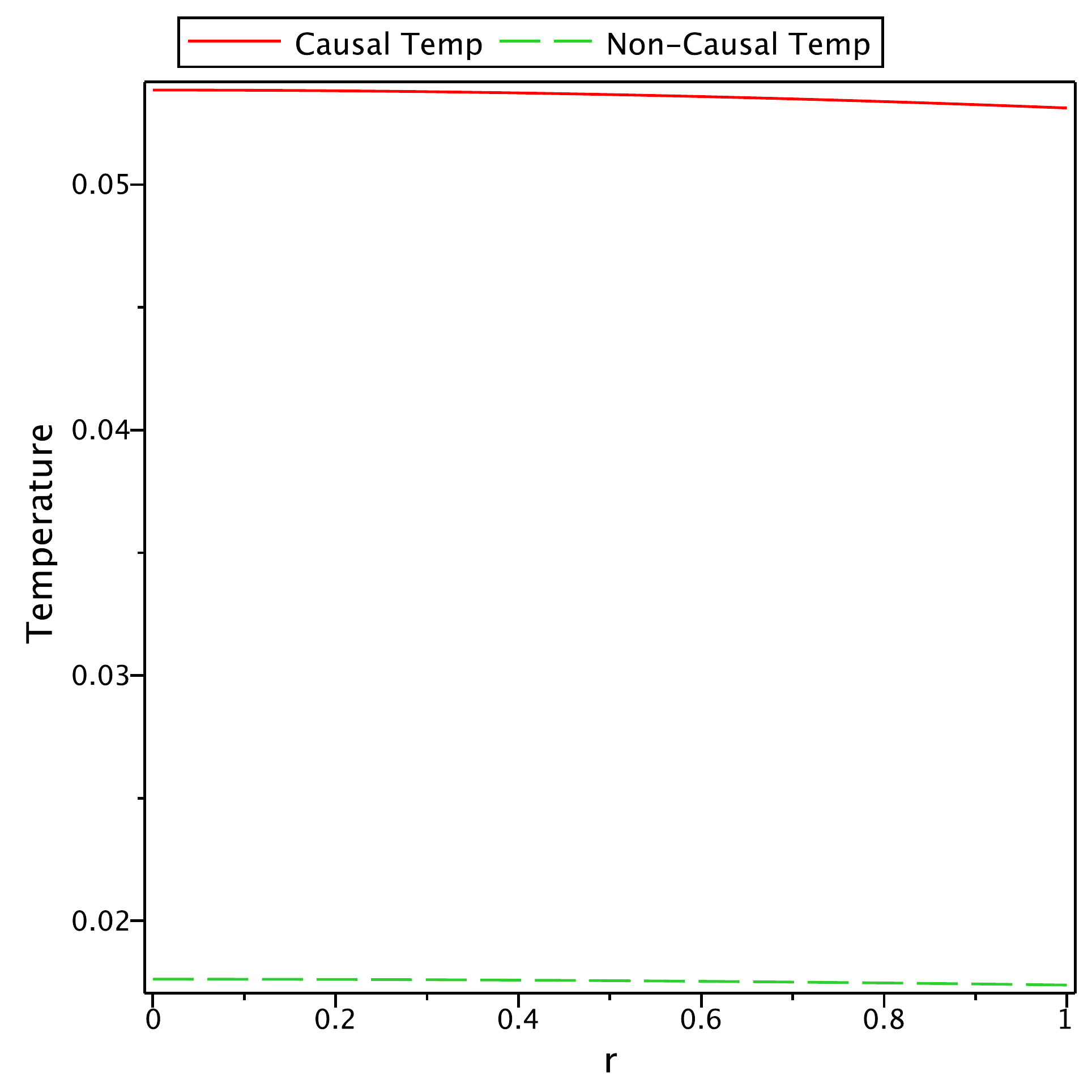}
		\caption{}
		\label{fig:Tempf2}
	\end{subfigure}
	\begin{subfigure}{0.333\textwidth}
		\centering
		\includegraphics[width=\linewidth]{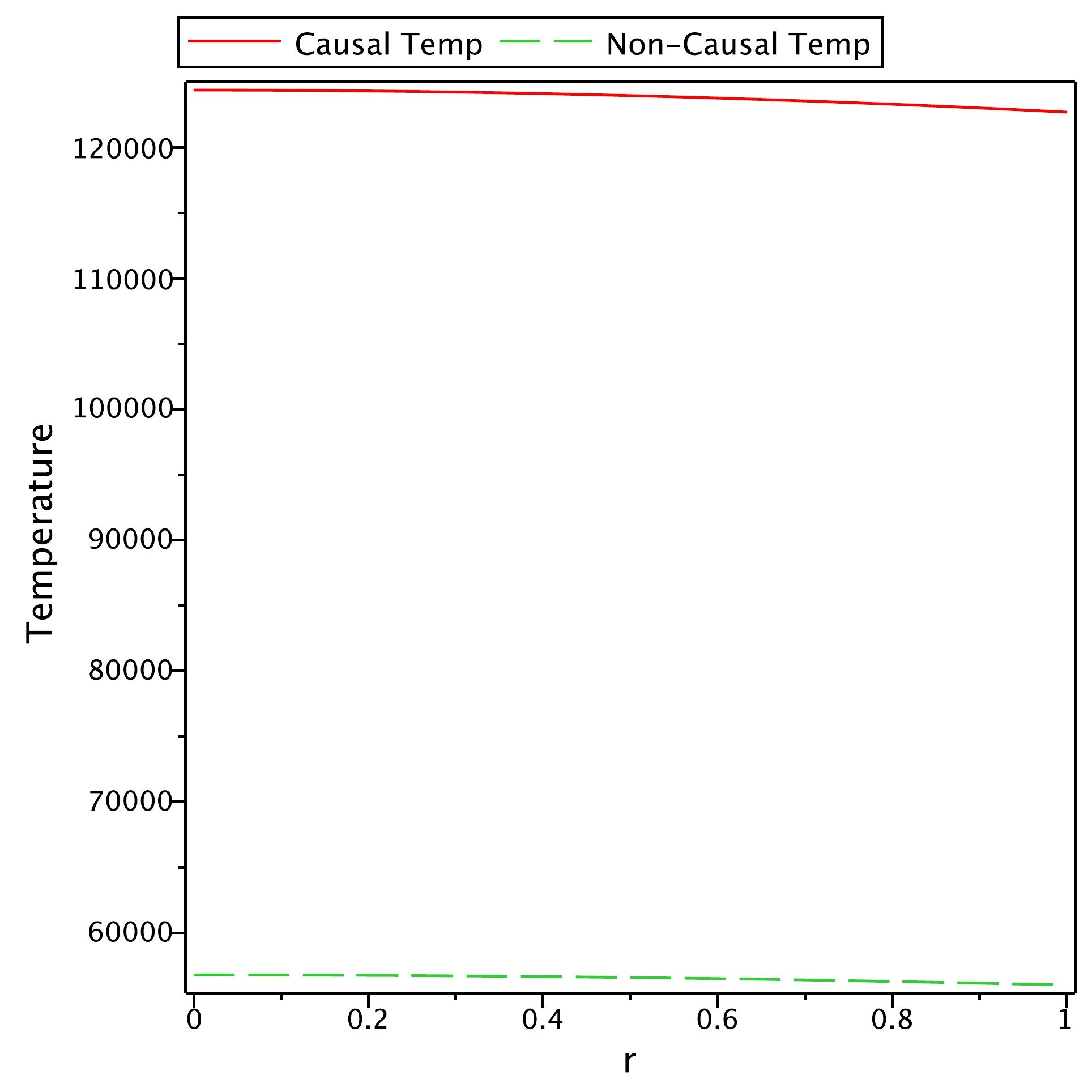}
		\caption{}
		\label{fig:Tempf3}
	\end{subfigure}
	\caption{(a), (b) \& (c) shows the plots of the temperature profiles of the collapsing stellar system w.r.t. radial coordinate $r$ for $\sigma=0$ and for $f(R)=R+\lambda R^{2}$ with $\lambda=5$, $f(R)=R^{1+\epsilon}$ with $\epsilon=0.01$ and $f(R)=R+\lambda \left[\exp\left(-\gamma R\right)-1\right]$ with $\lambda=0.1$ \& $\gamma=0.0002$ respectively.}
\end{figure}
Using conditions in equation \eqref{tkt},
the the causal heat transport equation \eqref{tempabf} becomes
\begin{eqnarray}
\beta T^{-\sigma} \left(qbs\right)_{,t}+q\,a\,b\,s&=&-\frac{\alpha \,\left(aT \right)_{,r}}{bs}\,\,T^{3-\sigma} .\label{tempabff}
\end{eqnarray}
The noncausal solution of the heat transport equation \eqref{tempabff},  with $\beta=0$ i.e. $\tau=0$ are\,\cite{GG2001}
\begin{eqnarray}
\left(a\,T\right)^{4}&=&-\frac{4}{\alpha}\int a^{4}\,q\,b^2\,s^2\,dr+G(t),
\hspace{1.6cm} \sigma= 0  \label{Nsigman4}\\
\ln\left(a\,T\right)&=&-\frac{1}{\alpha}\int q\,b^2\,s^2\,dr+G(t).\hspace{2cm} \sigma=4  \label{Nsigma4}
\end{eqnarray}
The causal solution of the above heat transport equation \eqref{tempabff} are\,\cite{GG2001}
\begin{eqnarray}
\left(a\,T\right)^{4}&=&-\frac{4}{\alpha}\left[\beta\int a^3\,b\,s(q\,b\,s)_{,t}\,dr
+\int a^{4}\,q\,b^2\,s^2\,dr\right]+G(t), \hspace{0.6cm} \sigma=0  \label{Csigman4}\\
\left(a\,T\right)^{4}&=&-\frac{4\beta}{\alpha}\exp\left(-\int\frac{4\,q\,b^{2}\,s^{2}}{\alpha}\,dr\right)
\int a^3\,b\,s(q\,b\,s)_{,t}\,dr\,\exp\left(\int\frac{4\,q\,b^{2}\,s^{2}}{\alpha}\,dr\right)\nonumber\\
&&+G(t)\exp\left(-\int\frac{4\,q\,b^{2}\,s^{2}}{\alpha}\,dr\right),\hspace{4.2cm} \sigma=4 \label{Csigma4}
\end{eqnarray}
where $G(t)$ appears as a function of integration and is determined by following boundary condition
\begin{eqnarray}
\left(T^{4} \right)_{\Sigma}&=&\left( \frac{L_{\infty}}{4\pi\delta r^{2}b^{2}s^{2}} \right)_{\Sigma}. \label{Tsigma}
\end{eqnarray}
where $L_{\infty}$ is the total luminosity for an observer at infinity given by \eqref{LInfinity}
and $\delta>0$ is constant. 
The Figs. \ref{fig:Temp}, \ref{fig:Tempf2} \& \ref{fig:Tempf3} shows that for all these three $f(R)$ models under consideration, the stellar system deviates from thermodynamical equilibrium due to the relaxation effects.
Also, the causal and noncausal temperature profiles differ inside the interior of the star and causal temperature remains greater than that of the non-causal temperature.
\section{Discussion of the results}\label{sec6}
In this paper, we investigated the dynamics of 
a collapsing matter configuration in $f(R)$ gravity.
The matter is assumed to be so massive that no stable compact objects
like a white dwarf or a neutron star forms. 
The interior spacetime is smoothly matched to 
the outgoing radiation Vaidya metric across a timelike hypersurface.
Incidentally, as has been noted earlier too, the
matching conditions for the $f(R)$ gravity is highly restrictive, since the
geometric variables which are to matched here
not only includes induced metric and the extrinsic curvatures,
but also the trace of the extrinsic curvatures, and the Ricci scalar along with
it's time derivative.
However, we have shown that all these matching conditions can be carried out consistently,
leading to a spacetime solution which admits a collapsing scenario in which the 
matter cloud radiates heat flux, in such a manner that the entire matter is radiated out 
without forming a black hole. Although similar solutions have been reported earlier
for GR, our solution incorporates these features into the collapsing models
of three particular $f(R)$ gravity models, while maintaining all the energy conditions.
In particular, for all the three $f(R)$ theories we have analyzed physical quantities like 
energy density, in eqn. \eqref{rhoabf}, radial pressure\,\eqref{prabf} and 
tangential pressure in eqn.\eqref{ptabf}, pressure anisotropy, in eqn. \eqref{Delta1} 
and it can be seen from the Figs. \ref{fig:rho11}, \ref{fig:rho12} \& \ref{fig:rho13}\,,
\ref{fig:pr11}, \ref{fig:pr12} \& \ref{fig:pr13}\,,
\ref{fig:pt11}, \ref{fig:pt12} \& \ref{fig:pt13}\,, 
\ref{fig:delta11}, \ref{fig:delta12} \& \ref{fig:delta13} that they are regular and positive throughout the collapse. 
From Figs.\,\ref{fig:q11}, \ref{fig:q12} \& \ref{fig:q13}
it is also clear that 
the radial heat flux\,\eqref{qabf} is finite and positive throughout collapse. 
In particular, for $f(R)=R+\lambda R^{2}$, Figs. \ref{fig:Lumf1} and \ref{fig:adf1}
show that both total luminosity and the effective adiabatic index are positive and
 increasing and have maximum value where luminosity is maximum. This behavior of the luminosity and 
 adiabatic index can be interpreted as follows: an observer at rest at infinity will see a exponential 
 radiating radial source until it reaches time when luminosity reaches its maximum value 
 and then instantaneous turn off of the radial source. This happens since
 the total mass of the star radiates linearly as seen from the Fig. \ref{fig:m},
  and when the star reaches its maximum luminosity, there is no mass left to radiate 
  and hence the observer at rest at infinity will see sudden turn off of the light source.
  The similar kind of behavior were obtained in GR too \cite{GP_RC}. The Fig. \ref{fig:adf1} 
  also shows that the effective adiabatic index is positive and less than $4/3$ which implies that the considered stellar system is unstable and represents a collapsing phenomena. Also note that
  the Figs. \ref{fig:E11}, \ref{fig:E12} \& \ref{fig:E13} show that the energy conditions
are positive and regular throughout 
the interior of the star.

For $f(R)=R^{1+\epsilon}$ and $f(R)=R+\lambda\,\left[\exp{(-\sigma R)}-1\right]$
similar behavior is obtained as well.
For example, Figs. \ref{fig:adf2}, and \ref{fig:adf3} shows that the effective adiabatic 
index is constant function of time, and is positive and less than $4/3$,
and hence represents a collapsing scenario. 
We have shown graphically that the star radiates all 
its mass before reaching the singularity. So, there are no trapped surfaces formed during the collapse.
This implies that neither a black hole nor a naked singularity exist at the end state of collapse. 
The Figs. \ref{fig:E21}, \ref{fig:E22} \& \ref{fig:E23} and \ref{fig:E31}, \ref{fig:E32} \& \ref{fig:E33} show that these model under consideration are
physically viable. 
Also, the results obtained here reduces to those for GR for $f(R)=R$ \cite{SCJ}.

Let us now comment on the nature of the central singularity. First, we note 
that the Ricci scalar \eqref{RScalar} together with \eqref{s(t)} imply that
it diverges at $t=0$, when all the matter has been radiated away. So, naturally the 
question arises regarding the gravitational strength of the central curvature singularity
(a strong singularity would imply a naked singularity). However, in the present case, 
the curvature scalars go as $t^{-2}$ which is precisely the sufficient condition
for the singularity to be weak. So, our solutions represent a physically viable model
where the spherically symmetric collapsing matter cloud undergoes
gravitational collapse, which, during the collapse, also radiates away mass in the form of heat flux. 
The flux is radiated at such a rate that no horizon is ever formed and the central
singularity is naked but gravitationally weak in nature. 
\\\\{\bf{Acknowledgments}}\\
The author AC is supported by the DST-MATRICS scheme of
government of India through MTR/2019/000916 and by the
DAE-BRNS Project No. 58/14/25/2019-BRNS.
\\\\{\bf{Data availability}}\\
This manuscript has no associated data. This is a theoretical
study and does not contain any experimental data.
%

\section*{Appendix}
In this appendix, we give the detail expressions of the physical quantities of the collapsing matter cloud in terms of the metric functions.
More precisely, we give the values for \eqref{rho}-\eqref{q}, and other
quantities like the expansion scalar \eqref{Theta} and the Misner- Sharp mass function \eqref{mass}.
\begin{eqnarray}
\rho&=&\frac{6F C_{3}C_{4}^{3}}{S_{1}C_{Z}^{2}t^{2}}\left[C_{1}
-2C_{2}C_{3}\left( 2C_{4}+C_{3}r^2\right)\right]^{2}+\frac{3\dot{s}\dot{F}}{s\,a^{2}}-\frac{F^{\prime\prime}}{b^{2}\,s^{2}}+\frac{f-R\,F}{2}
-\frac{F^{\prime}}{b^{2}\,s^{2}}\left[\frac{b^{\prime}}{b}+\frac{2}{r}\right],\label{rhoabf}\\
p_{r}&=&\frac{F C_{3}C_{4}^{2}}{S_{1}C_{Z}^2t^{2}}\left[2C_{1}C_{2}C_{3}\left(12C_{4}^{2}+4C_{3}C_{4}r^2-C_{3}r^{2}\right)-C_{1}^{2}\left(4C_{4}-C_{3}r^2\right)-8C_{2}^{2}C_{3}^2C_{4}\left(2C_{4}+C_{3}r^2\right)^2 \right]\nonumber\\
&&~~~~~~~~~~~~~~~~~~~~~~~~~~~~~~~~~~~~-\left(\frac{f-R\,F}{2}\right)-\frac{\dot{F}}{a^2}\left(\frac{\ddot{F}}{\dot{F}}+\frac{2\dot{s}}{s}\right)
+\frac{F^{\prime}}{b^{2}\,s^{2}}\left(\frac{a^{\prime}}{a}+\frac{2}{r}+\frac{2b^{\prime}}{b} \right),\label{prabf}\\
p_{t}&=&\frac{F C_{3}C_{4}^{2}}{S_{1}C_{Z}^2t^2}\left[C_{1}^{4} C_{4}^{2} \left(C_{3} r^{2}-4 C_{4}\right)
+2 C_{1}^{3} C_{2} C_{3} C_{4}^2 \left(-3 C_{3}^{2} r^{4}+8 C_{3} C_{4} r^{2}+28 C_{4}^{2}\right)\right]+\frac{F^{\prime}}{r\,b^{2}\,s^{2}}\nonumber\\
&+&\frac{2F C_{1}^{2} C_{3}C_{4}^{2}}{S_{1}C_{Z}^2t^2}\left[ \left(C_{3} r^{2}+2 C_{4}\right)^{2} \left[6 C_{2}^{2} C_{3}^{2} C_{4}^{2} 
\left(C_{3} r^{2}-6 C_{4}\right)+C_{Z}^{2} \left(2 C_{4}-C_{3} r^{2}\right)\right]\right]+\frac{F^{\prime} a^{\prime}}{a\,b^{2}\,s^{2}}\nonumber\\
&+&\frac{2 F C_{1} C_{2}C_{3}^2C_{4}^{2}}{S_{1}C_{Z}^2t^2}\left[\left(C_{3} r^2+2 C_{4}\right)^{3} \left[C_{Z}^{2} 
\left(C_{3} r^{2}-6 C_{4}\right)-4 C_{2}^{2} C_{3}^{2} C_{4}^{2} \left(C_{3} r^{2}-10C_{4}\right)\right]\right]+\frac{F^{\prime\prime}}{b^{2}\,s^{2}} \nonumber\\
&-&\frac{8F C_{2}^{2} C_{3}^{3}C_{4}^{3}}{S_{1}C_{Z}^2t^2}\left[ \left(C_{3} r^{2}+2 C_{4}\right)^4 
\left(4C_{2}^{2} C_{3}^{2}C_{4}^{2}-C_{Z}^{2}\right) \right]-\left(\frac{f-R\,F}{2}\right)-\frac{\dot{F}}{a^2}\left(\frac{\ddot{F}}{\dot{F}}+\frac{2\dot{s}}{s}\right),\label{ptabf}\\
q&=&-\left[\frac{ C_{4}^{5/2} C_{3}^{3/2}}{C_{Z}^{2} t^{3}}\frac{4 \sqrt{2}r F C_{1}\left[4 C_{2} C_{3}
	-\frac{2 C_{1}}{C_{3} r^{2}+2 C_{4}}\right]}{ \left[C_{4}^{2} 
	\left(4 C_{2} C_{3}-\frac{2 C_{1}}{C_{3} r^{2}+2 C_{4}}\right)^{2}-4 C_{Z}^{2}\right]^{3/2}}\right]+\frac{1}{a^{2}\,b^{2}\,s^{2}} \left[\dot{F}\,^{\prime}-\frac{\dot{F}a^{\prime}}{a}-\frac{\dot{s}\,F^{\prime}}{s} \right],\label{qabf}\\
\Theta &=& \frac{6\sqrt{C_{3} C_{4}}}{t \sqrt{C_{4}^{2} \left(2 C_{2} C_{3}
		-\frac{C_{1}}{C_{3} r^{2}+2 C_{4}}\right)-2 C_{Z}^{2}}},\label{Thetaabf}\\
m&=&\frac{8 t r^{3} C_{3} C_{4}^3 C_{Z}}{\left(C_{3} r^{2}+2 C_{4}\right)^{3}}
\,\left[\frac{2 C_{2} C_{3} \left(C_{3} r^{2}+2 C_{4}\right)-C_{1}}
{2 \left(C_{3} r^{2}+2 C_{4}\right) \left(C_{2} C_{3} C_{4}^{2}-C_{Z}^{2}\right)-C_{1} C_{4}^{2}}\right],\label{mabf}\\
S_{1}&=&C_{4}^{2}\left(2\,C_{2}\,C_{3}-{C_{1}}/\left({2\,C_{4}+C_{3}\,r^2}\right)\right)^{2}-C_{Z}^{2}.\nonumber
\end{eqnarray}
\end{document}